\documentclass[sigplan,10pt]{acmart}
\AtBeginDocument{%
  }
\setcopyright{acmlicensed}
\copyrightyear{2018}
\acmYear{2018}
\acmDOI{XXXXXXX.XXXXXXX}
\acmConference[Conference acronym 'XX]{Make sure to enter the correct
  conference title from your rights confirmation email}{June 03--05,
  2018}{Woodstock, NY}
\acmISBN{978-1-4503-XXXX-X/2018/06}

\usepackage{algorithmicx}
\usepackage[ruled]{algorithm}
\usepackage{algpseudocode}

\usepackage[T1]{fontenc}
\usepackage{eurosym}
\usepackage[only,bigsqcap,bigparallel,fatsemi,interleave,sslash]{stmaryrd}
\usepackage{textcomp}
\usepackage{isabelle,isabellesym}

\isabellestyle{it}
\allowdisplaybreaks[3]

\newcommand{\prob}{\mathcal{P}}

\newcommand{\bind}{\mathbin{{>}\!\!\!{>}\!\!{=}}}
\newcommand{\return}{\mathop{\mathrm{return}}}

\newcommand{\inverse}{^{-1}}
\newcommand{\Nat}{\mathbb{N}}

\newcommand{\Real}{\mathbb{R}}

\newcommand{\DP}[2]{\mathrm{DP}^{#1}_{#2}}

\begin{document}
\title{Formalization of R\'{e}nyi Differential Privacy in Isabelle/HOL (Draft version)} 
\author{Tetsuya Sato}
\correspondingauthor
\authornotemark[1]
\email{sato\_tetsuya@iwate-pu.ac.jp}
\orcid{0000-0001-9895-9209}
\affiliation{%
  \institution{Iwate Prefectural University}
  \city{Sugo, Takizawa}
  \state{Iwate}
  \country{Japan}
}

\renewcommand{\shortauthors}{T. Sato}

\begin{abstract}
Differential privacy (DP) is a gold-standard definition of privacy designed to evaluate the privacy guarantee of databases.
R\'{e}nyi differential privacy (RDP) is an important variant of DP that gives tight evaluations of machine learning algorithms.
These formulations are easy to understand, but evaluating privacy for actual databases is complicated.
Even a small change in the database can break its privacy, and evaluations of DP and RDP themselves are complicated.
In particular, RDP evaluations in recent studies have become much harder.
Therefore, formal verification of them has been studied for over a decade.

In this paper, we propose an Isabelle/HOL library for formalizing R\'{e}nyi differential privacy in a general setting.
We first formalize RDP for continuous distributions.
As a technical basis, we formalize several properties of convex functions and $f$-divergences for nonnegative convex functions.
We formalize basic properties of RDP, conversion laws from RDP to DP, the moment accountants in DP, and the Gaussian mechanism and its RDP.
Finally, we discuss the stochastic gradient descent mechanism with Gaussian noise and formalize a rough evaluation of its RDP and DP.
\end{abstract}


\received{20 February 2007}
\received[revised]{12 March 2009}
\received[accepted]{5 June 2009}

\maketitle

\section{Introduction}
Differential privacy(DP) is a de facto standard definition of privacy for databases with large amounts of individuals' private data~\cite{Dwork2006,DworkMcSherryNissimSmith2006}.
It is already applied in the real world, such as iOS, Google Maps, and the United States census~\cite{AppleDP, GoogleDP, Gong2022Harnessing}.
As notable strengths of DP, it guarantees robustness against any kind of statistical attack, and its composition theorems allow decomposing the evaluation of the privacy guarantee of complex databases into their components' ones.

In the study of machine learning algorithms, privacy-preserving learning algorithms with noise sampled from Gaussian distributions have attracted both industry and academia.
To evaluate DP of such \emph{Gaussian mechanisms} rigorously, variants of differential privacy based on R\'{e}nyi divergence have become increasingly important.
Abadi et al. introduced gradient descent algorithms with Gaussian noise and discussed their moment accountants, which can be captured with R\'{e}nyi divergences~\cite{AbadiGaussMomentAccountant};
Bun et al. introduced zero-concentrated differential privacy~\cite{BunS16};
Mironov introduced R\'{e}nyi differential privacy(RDP)~\cite{Mironov17}.
There are many studies on DP and RDP mechanisms.

In general, evaluating DP and RDP for databases is cumbersome.
A few, even one-line, changes to database programs can break their differential privacy~\cite[Sec. 3]{10.14778/3055330.3055331}.
In addition, RDP is much more complicated than DP, and evaluating it is not obvious, even understanding the proofs (For example, the original proof of Lemma \ref{convRDPtoDP:Asoodeh} is complicated).
Formal verification of RDP is more important and more complex than that of DP.
For these reasons, formal verification of DP and RDP has been studied for over a decade.
The first approach is based on Benton's relational Hoare logic~\cite{Benton2004export:67345}.
Starting from Barthe et al.'s relational program logic for reasoning about DP~\cite{Barthe:2012:PRR:2103656.2103670}, many variants are studied~\cite{Barthe:2015:HAR:2676726.2677000,Barthe:2016:APC:2976749.2978391,2016arXiv160105047B}, and extended to support RDP and more general divergences~\cite{BartheOlmedo2013,DBLP:conf/lics/SatoBGHK19,Sato_Katsumata_2023}.

The second is reformulating DP as the sensitivity of functions between metric spaces.
Based on this approach, Gaboardi et al. introduced a linear dependent type system for reasoning about differential privacy~\cite{GHHNP13,10.1145/2746325.2746335}.
It was applied for automated reasoning~\cite{10.1145/3589207} and verification of integrity~\cite{10.1145/2741948.2741978}.

The third approach is formalizing DP and RDP directly in \emph{proof assistants}.
Tristan et al. proposed the library {\tt SampCert} for verified implementations of DP and RDP using Lean~\cite{10.1145/3729294,Tristan_SampCert_Verified_2024}.
It contains the formalization of DP and RDP in the \emph{discrete setting} and the discrete-Laplace and discrete-Gaussian mechanisms (see also \cite{canonne2021discretegaussiandifferentialprivacy}).
Feng formalized functions of the OpenDP library~\cite{verification_opendp_transformations} in Lean.

\paragraph{Motivation}
To our knowledge, RDP based on the continuous model of probabilistic programs has not been formalized yet.
In this paper, we propose an Isabelle/HOL library for semantic-level formal verification of RDPs supporting \emph{continuous} probability distributions, including Gaussian distributions.

First, the continuous setting is more general than the discrete one.
Moreover, Isabelle/HOL's discrete probability distributions are implemented as measures on discrete spaces.

Second, even in the study of differential privacy in the discrete setting, we might need continuous models.
For example, Mironov uses the continuous ``ideal mechanism'' to verify a secure floating-point implementation\footnote{It is different from discrete-Laplace mechanism given in \cite{canonne2021discretegaussiandifferentialprivacy}} of the Laplace mechanism~\cite{Mironov2012OnSO}.
In addition, many pencil-and-paper proofs on RDP are written in the continuous setting.
It is natural and convenient for us to implement them directly.

Finally, the author thought that it must not be so hard to formalize it in the continuous setting.
We later formalize that $f$-divergences (hence RDP) in the continuous setting can be approximated with ones in the discrete setting.
\paragraph{Contribution}
In this paper, we give an Isabelle/HOL library for R\'{e}nyi differential privacy in the continuous setting, and give a formalization of the stochastic gradient descent (SGD) mechanism with Gaussian noise and Poisson subsampling.

For a technical basis, we formalize $f$-divergence, which is a class of statistical divergences generated by convex functions.
We first give formal proofs of several basic properties of convex real functions, such as left and right derivatives and piecewise monotonicity.
Then we formalize $f$-divergence for nonnegative generator functions, and its basic properties such as postprocessing inequality and joint convexity.
Furthermore, we formalize an approximation property of $f$-divergences called approximability, which is a technical basis of the composition theorem of R\'{e}nyi differential privacy in the general setting.

We then formalize basic properties of RDP, conversion laws from RDP to DP, the moment accountants in DP, and the Gaussian mechanism and its RDP.
Finally, we formalize a rough evaluation of RDP and DP of Gaussian SGD.
\section{Preliminaries}
Throughout this paper, we write $\Real$ and $\Nat$ for the sets of real numbers and natural numbers, respectively.
We write $(0,\infty)$, $[0,\infty)$, and $[0,\infty]$ for the sets of \emph{positive} reals, and \emph{nonnegative} reals and nonnegative \emph{extended} reals respectively.

\subsection{Measure theory}
A \emph{measurable space} $X = (|X|,\Sigma_X)$ is a set $|X|$ with a $\sigma$-algebra $\Sigma_X$ over $|X|$ which is a nonempty family of subsets of $|X|$ closed under countable unions and complelemts.
For a measurable space $X$,  we write $|X|$ and $\Sigma_X$ for the underlying set and $\sigma$-algebra, respectively.
For simplicity, we often write just $X$ for the underlying set $|X|$.
For example, $x \in X$ means $x \in |X|$.
As an abuse of notation, we write $\Real$ for the measurable spaces with the Borel algebra, the least $\sigma$-algebra containing all open intervals. 
A \emph{measurable function} $f \colon X \to Y$ is a function $f \colon |X| \to |Y|$ satisfying $f\inverse(S) \in \Sigma_X$ for each $S \in \Sigma_X$.
A \emph{measure} $\mu$ on a measurable space $X$ is a function $\mu \colon \Sigma_X \to [0,\infty]$ having the countable additivity.
A measure $\mu$ on $X$ is called a \emph{probability measure} if $\mu(X) = 1$ respectively.
We define the sum $\mu_1 + \mu_2$ of two measures $\mu_1,\mu_2$ on $X$ by $(\mu_1 + \mu_2)(S) = \mu_1(S) + \mu_2(S)$,
and the scaling $a \mu$ of measure $\mu$ on $X$ with $0 \leq a$ by $a\mu(S) = a \dot \mu(S)$.

\paragraph{Giry monad}
For a measurable space $X$, we define the measurable space $\prob (X)$ as follows:
$|\prob (X)|$ is the set of probability measures on $X$, and 
$\Sigma_{\prob (X)}$ is the coarsest $\sigma$-algebra making 
each evaluation mappling $\mu \mapsto \mu(A)$ ($A \in \Sigma_X$) a measurable function of type $\prob(X) \to [0,\infty]$. 
The constructor $\prob$ forms a monad $(\prob,\return,\bind)$ called the \emph{Giry monad}\cite{Giry1982}.
The unit $\return$ is defined by $\return (x) = \mathbf{d}_x$ where $\mathbf{d}_x$ is the Dirac measure on $X$ centered at $x \in X$\footnote{It is defined by 
$\mathbf{d}_x (A) = 1$ if $x \in S$ and $\mathbf{d}_x (A) = 0$ otherwise}.
For any $f\colon X \to  \prob(Y)$ and $\mu \in \prob (X)$, $(\mu \bind f) \in \prob (Y)$ is defined by $(\mu \bind f)(A) = \int f(x)(A)~d\mu(x)$.
We can interpret a probabilistic program as a measurable function $c \colon X \to \prob(Y)$.
The sequential composition of $c \colon X \to \prob(Y)$ and $c' \colon Y \to \prob(Z)$ can be interpreted by $(\lambda x. c(x) \bind c')$.

\paragraph{Do notation}
For better readability, 
we introduce the syntactic sugars called \emph{do notation}:
\begin{align*}
\{x \leftarrow \mu; f(x) \} &= (\mu \bind f)\\
\{x \leftarrow \mu_1; y \leftarrow \mu_2(x); ... \} &= \{x \leftarrow \mu_1; \{ y \leftarrow \mu_2(x); ... \} \}
\end{align*}
We notice that the Giry monad is commutative.
Using do notation, the commutativity is written as follows:
for $\mu_1 \in \prob(X)$, $\mu_2 \in \prob(Y)$, and measurable $f \colon X \times Y \to \prob(Z)$,
\[\textstyle
\{x \leftarrow \mu_1; y \leftarrow \mu_2; f(x,y) \}= \{y \leftarrow \mu_2;x \leftarrow \mu_1; f(x,y) \}.
\]

\subsection{Measure theory in Isabelle/HOL}
The Isabelle/HOL's standard library contains the formalizations of measure theory and the Giry monad.

The type \isa{{\isacharprime}a measure} is the type of \emph{measure space}, that is, the triple $(|X|,\Sigma_X,\mu)$ consisting of a measurable space $(|X|,\Sigma_X)$ and a measure $\mu$ on it.
The standard library provides the projections of these components:
\begin{align*}
\isa{space\ }&\isa{:: {\isacharprime}a measure {\isasymRightarrow} {\isacharprime}a set } \\
\isa{sets\ }&\isa{:: {\isacharprime}a measure {\isasymRightarrow} {\isacharprime}a set set }\\
\isa{emeasure\ }&\isa{:: {\isacharprime}a measure {\isasymRightarrow} {\isacharprime}a set {\isasymRightarrow} ennreal}
\end{align*}
Here, \isa{ennreal} is the type for $[0,\infty]$, and \isa{{\isacharprime}a set } is the type of all subsets with elements of type \isa{{\isacharprime}a}.

The type \isa{{\isacharprime}a measure} is used for both measurable spaces and measures.
We sometimes need to distinguish them. 
For example, \isa{borel::real\ measure} is used for the Borel space $\Real$, and 
\isa{lborel::real\ measure} is for the Lebesgue measure\footnote{Precisely, the Lebesgue measure is \isa{completion\ lborel}.} on $\Real$.

The set of measurable functions from \isa{X} to \isa{Y} is given by \isa{X\ {\isasymrightarrow}\isactrlsub M\ Y}, hence measurability of \isa{f} is denoted by \isa{f\ {\isasymin}\ X\ {\isasymrightarrow}\isactrlsub M\ Y}.

For a measure \isa{M :: {\isacharprime}a measure} and \isa{f\ {\isasymin}\ M\ {\isasymrightarrow}\isactrlsub M (borel::real\ measure)}
\footnote{Here, \isa{M} is also used for the underlying measurable space on \isa{M}.},
the integration of \isa{f} with measure \isa{M} (over a set \isa{A}) is denoted by \isa{{\isasymintegral}x{\isachardot}{\kern0pt}\ f\ x\ {\isasympartial}M} (resp. \isa{{\isasymintegral}x{\isasymin}A{\isachardot}{\kern0pt}\ f\ x\ {\isasympartial}M}).

Isabelle/HOL's standard library contains the formalization of the Giry monad and its commutativity.
The components in $(\prob,\return,\bind)$ are provided as the following constants. 
\begin{align*}
\isa{prob\_algebra\ }\isa{:: {\isacharprime}a measure {\isasymRightarrow}}&\isa{ {\isacharprime}a measure measure }\\
\isa{return\ }\isa{:: {\isacharprime}a measure {\isasymRightarrow}}&\isa{ {\isacharprime}a {\isasymRightarrow} {\isacharprime}a measure}
\end{align*}
\vspace{-1.5\baselineskip}
\[
\isa{{\isasymbind}\ }\isa{:: {\isacharprime}a measure {\isasymRightarrow}}\isa{ {\isacharparenleft}{\isacharprime}a {\isasymRightarrow} {\isacharprime} b measure {\isacharparenright} {\isasymRightarrow} {\isacharprime}b measure}
\]
\section{Differential Privacy}
\emph{Differential privacy}(DP) is a statistical definition of privacy introduced by Dwork et al.\cite{Dwork2006,DworkMcSherryNissimSmith2006}.
Later, Mironov introduced \emph{R\'{e}nyi differential privacy}(RDP)~\cite{Mironov17} in the line of studies about moment accountants in DP to make tight privacy guarantees for Gaussian mechanisms~\cite{AbadiGaussMomentAccountant,BunS16}.
In this work, we use both kinds of differential privacy.
In this section, we give a crash course on differential privacy and the prior study for the formalization of DP in Isabelle/HOL~\cite{10.1145/3703595.3705875,Differential_Privacy-AFP}.

A basic approach to differential privacy is to prevent distinguishing between a dataset and the dataset plus an individual's data entry $x$.
Thus, differential privacy is based on the \emph{adjacency} of datasets.
For a basic example, $D_1$ and $D_2$ are adjacent if $D_2$ is made by insertion of an entry $x$ to $D_1$ (i.e. $D_2 = D_1 \cup \{x\}$).
For another example, two datasets $D_1$ and $D_2$ are adjacent if they are different only in one record, namely, their Hamming distance is $1$ (i.e. $d_H(D_1,D_2) = 1$).
Moreover, for datasets of user-related events, we can discuss two kinds of differential privacy: user-level and event-level~(see \cite[Sec. 1]{10.1145/1806689.1806787}).
Generally speaking, adjacency can be captured by an adjacency relation $R_{\mathrm{adj}}$ between datasets.

Differential privacy assumes queries with noise-adding anonymization and is designed to measure the indistinguishability of noisy outputs from two adjacent datasets.
Hence, we formulate a noisy database mechanism as a measurable function $c \colon X \to \prob(Y)$ where $X$ and $Y$ are the domains of inputs and outputs, respectively.
We now give the standard definition of differential privacy (cf. \cite{10.1007/11761679_29}).
\begin{definition}[{Differential Privacy(DP)~\cite[Def. 2.4]{DworkRothTCS-042}}]\label{def:DP}
Let $R_{\mathrm{adj}} \subseteq X \times X$ be an adjacency relation, and let $0 \leq \varepsilon$ and  $0 \leq \delta$.
A database mechanism $c \colon X \to \prob(Y)$ is $(\varepsilon,\delta)$-differentially private(DP)
if for two adjacent datasets $(D_1,D_2) \in R_{\mathrm{adj}}$, 
\[
\forall S \in \Sigma_Y.~  \Pr[S \in c(D_1)] \leq \exp(\varepsilon) \Pr[S \in c(D_2)] + \delta.
\]
\end{definition}
Intuitively, $\varepsilon$ is an upper bound of probability ratio between $c(D_1)$ and $c(D_2)$, and $\delta$ is an error.
When $\varepsilon$ and $\delta$ are small, $c(D_1)$ and $c(D_2)$ are hard to distinguish\footnote{In particular, if $(\varepsilon,\delta) = (0,0)$ then $c(D_1) = c(D_2)$.}.

A key feature of DP is the \emph{composition theorems}, which are a compatibility condition for compositions.
For example, we have the following proposition theorem.
\begin{proposition}[{Adaptive composition~\cite[Thm. B.1]{DworkRothTCS-042}}]
If $c_1 \colon X \to \prob(Y)$ is $(\varepsilon_1,\delta_1)$-DP and 
$c_2 \colon X \times Y \to \prob(Z)$ satisfies that 
for any $y \in Y$,  $(\lambda x. c_2(x,y)) \colon X \to \prob(Y)$ is $(\varepsilon_2,\delta_2)$-DP.
then $(\lambda x. \{y\leftarrow c_1(x); c_2(x,y) \})$ is $(\varepsilon_1+\varepsilon_2,\delta_1+\delta_2)$-DP.
\end{proposition}

\paragraph{Divergence for DP}
The indistinguishability in DP can be reformulated equvalently by a \emph{statistical divergence}, which is a ``distance'' between probability distributions.
We recall $\DP \varepsilon X \colon |\prob(X)| \times |\prob(X)| \to [0,\infty)$ showed in \cite{BartheOlmedo2013,DBLP:conf/lics/SatoBGHK19,Sato_Katsumata_2023}: 
\[
\DP \varepsilon X (\mu_1,\mu_2) = \max_{S \in \Sigma_X}(\mu_1(S) - \exp(\varepsilon) \mu_2(S)).
\]
We then have that $c \colon X \to \prob(Y)$ is $(\varepsilon,\delta)$-DP \emph{if and only if}
$\DP \varepsilon X (c(D_1),c(D_2)) \leq \delta$ holds for all $(D_1,D_2) \in R_{\mathrm{adj}}$.
The above composition theorem and the stability for post-processes are derived from the below composability:
\begin{proposition}[Composability]
Let $\mu_1, \mu_2 \in \prob(X)$ and $c_1, c_2 \colon X \to \prob(Y)$.
If $\DP {\varepsilon_1} X(\mu_1,\mu_2) \leq \delta_1$ and $\DP {\varepsilon_2} Y (c_1(x),c_2(x)) \leq \delta_2$ for any $x \in X$
then $\DP { \varepsilon_1 +  \varepsilon_2} Y (\mu_1 \bind c_1, \mu_2 \bind c_2) \leq \delta_1 + \delta_2$.
\end{proposition}

\subsection{R\'{e}nyi differential privacy}
Noise-adding mechanisms with Gaussian distributions (called \emph{Gaussian mechanism}) are an essential component of privacy-preserving data analysis.
However, we cannot avoid errors in its DP evaluations.
The Gaussian distribution $\mathcal{N}(m,\sigma^2)$ with average $m$ and variance $\sigma^2$is given by the following probability density function
\[
f_{\mathcal{N}(m,\sigma^2)}(x) = \frac{1}{\sqrt{2\pi\sigma^2}} \exp \left(-\frac{(x-m)^2}{2\sigma^2}\right).
\]
The noise adding mechanism with $mathcal{N}(0,\sigma^2)$ can be written by
$(\lambda x. \{z \leftarrow \mathcal{N}(0,\sigma^2); \return (x+z) \})$.
By the shift property of Gaussian distributions, it is $(\lambda x. \mathcal{N}(x,\sigma^2))$.
For $x \neq y$, we evaluate the indistinguishability of $\mathcal{N}(x,\sigma^2)$ and $\mathcal{N}(y,\sigma^2)$ for $(\varepsilon,\delta)$-DP.
If $\delta = 0$ then $\varepsilon$ must be $\infty$, where the ratio between $f_{\mathcal{N}(y,\sigma^2)}(z)$ and $f_{\mathcal{N}(x,\sigma^2)}(z)$ is unbounded.

\emph{R\'{e}nyi differential privacy}(RDP) is a variant of differential privacy introduced by Mironov~\cite{Mironov17}.
Replacing the divergence $\DP \varepsilon {}$ for DP with the R\'{e}nyi divergence (of order $\alpha$) defined as follows, we obtain RDP.
\[
D^\alpha_X(\mu_1,\mu_2) = \frac{1}{\alpha - 1} \log \int \left( \frac{p_1(x)}{p_2(x)} \right)^\alpha p_2(x) ~d\pi(x)\quad (1 < \alpha).
\]
Here, $p_1(x)$ and $p_2(x)$ are density functions of $\mu_1$ and $\mu_2$ with respect to a base measure $\pi$.
\begin{definition}[{R\'{e}nyi Differential privacy(RDP)~\cite[Def. 4]{Mironov17}}]
Let $R_{\mathrm{adj}} \subseteq X \times X$ be an adjacency relation, and let $1 < \alpha$ and $0 \leq \rho$.
A database mechanism $c \colon X \to \prob(Y)$ is $(\alpha,\rho)$-R\'{e}nyi differentially private(RDP)
if $D^\alpha_X(c(D_1),c(D_2))$ holds for any adjacent datasets $(D_1,D_2) \in R_{\mathrm{adj}}$.
\end{definition}
R\'{e}nyi divergence satisfies $D^\alpha_X(\mu,\mu) = 0$ for all $\mu \in \prob(X)$, and has the composability.
\begin{lemma}[{Composability~\cite[Thm. 7]{DBLP:conf/lics/SatoBGHK19}}]
\label{Renyi:compose:monadic}
Let $\mu_1, \mu_2 \in \prob(X)$ and $c_1, c_2 \colon X \to \prob(Y)$.
If $D^\alpha_X(\mu_1,\mu_2) \leq \rho_1$ and for any $x \in X$, $D^\alpha_Y (c_1(x),c_2(x)) \leq \rho_2$ 
then $D^\alpha_Y (\mu_1 \bind c_1, \mu_2 \bind c_2) \leq \rho_1 + \rho_2$.
\end{lemma}
Postprocessing inequality $D^\alpha_Y (\mu_1 \bind c, \mu_2 \bind c) \leq D^\alpha_X(\mu_1,\mu_2)$ is just an application of this lemma for $c_1 = c_2$ and $\rho_2 = 0$.

The R\'{e}nyi divergence of Gaussian distributions is simple.
\begin{lemma}[{\cite[Prop. 7]{Mironov17}, modified}]
\label{lem:Renyi:Gaussian}
For $1 < \alpha$ and $0 < \sigma$, 
\[
D^\alpha_X(\mathcal{N}(x,\sigma^2),\mathcal{N}(y,\sigma^2))
= \frac{\alpha(x-y)^2}{2\sigma^2}.
\]
\end{lemma}
\paragraph{Basic properties of RDP}
From these facts, we obtain basic properties for RDP.
We have the composition theorem and the stability for postprocessing.
\begin{proposition}[{Composition theorem~\cite[Prop. 1]{Mironov17}}]
\label{RDP:compose:adaptive}
If $c_1 \colon X \to \prob(Y)$ is $(\alpha,\rho_1)$-RDP and 
$c_2 \colon X \times Y \to \prob(Z)$ satisfies that 
for any $y \in Y$,  $(\lambda x. c_2(x,y)) \colon X \to \prob(Y)$ is $(\alpha,\rho_2)$-RDP.
then $(\lambda x. \{y\leftarrow c_1(x); c_2(x,y) \})$ is $(\alpha,\rho_1+\rho_2)$-RDP.
\end{proposition}
\begin{proposition}[Postprocessing]
\label{RDP:postprocess}
If $c \colon X \to \prob(Y)$ is $(\alpha,\rho)$-RDP then
for any $g \colon Y \to \prob(Z)$, the composition
$(\lambda x. \{y \leftarrow c(x); g(y)\})  \colon X \to \prob(Z)$ is $(\alpha,\rho)$-RDP.
\end{proposition}
We also state the RDP of Gaussian mechanism.
\begin{lemma}
Let $1 < \alpha$ and $0 < \sigma$, and $f \colon X \to \Real$.
The Gaussian mechanism $(\lambda x. \mathcal{N}(x,\sigma^2)) \colon \Real \to \prob (\Real)$ is $(\alpha,\frac{\alpha r^2}{2\sigma^2})$-RDP, 
where $r = \sup\{  \| f(D_1) - f(D_2) \|_2~|~(D_1,D_2) \in R_{\mathrm{adj}}\}$.
\end{lemma}
This can be extended easily to multi-dimensional Gaussian mechanism $(\lambda \vec{x}. \mathcal{N}(\vec{x},\sigma^2 I )) \colon \Real^n \to \prob (\Real^n)$.
It is $n$ applications of 1-dimensional Gaussian mechanism $x_i \mapsto \mathcal{N}(x_i,\sigma^2)$.

\paragraph{Conversions from RDP to DP}
We can convert RDP to DP, comparing the divergences $\DP \varepsilon X(\mu_1,\mu_2)$ and $D^\alpha_X(\mu_1,\mu_2)$.
We give formal proofs of the following conversion laws.
\begin{lemma}[{\cite[Prop. 3]{Mironov17}}] 
\label{convRDPtoDP:Mironov}
Assume that $c$ is $(\alpha,\rho)$-RDP.
For $0 < \delta < 1$, $c$ is $\left(\rho - \frac{\log \delta}{\alpha-1},\delta\right)$-DP.
\end{lemma}
\begin{lemma}[{\cite[Thm. 20]{DBLP:journals/corr/abs-1905-09982}, fixed in \cite{DBLP:journals/corr/abs-2008-06529}}] 
\label{convRDPtoDP:Balle}
If $c$ is $(\alpha,\rho)$-RDP then it is $\left(\left((\rho - \frac{\log \delta + \log \alpha}{\alpha-1}+ \log\left(\frac{\alpha-1}{\alpha}\right)\right)_{+}\!\!,\delta\right)$-DP for $0 < \delta < 1$.
\end{lemma}
Here, $(\ldots)_{+}$ is the nonnegative part $\max (0,\ldots)$.
The term $\rho - \frac{\log \delta + \log \alpha}{\alpha-1}+ \log\left(\frac{\alpha-1}{\alpha}\right)$ can be negative, but in that case, we can set $\varepsilon = 0$. This conversion is known as a standard one. 
\begin{lemma}[{\cite[Lemma 1]{DBLP:journals/corr/abs-2008-06529}}] 
\label{convRDPtoDP:Asoodeh}
Assume that $c$ is $(\alpha,\rho)$-RDP.
For $0 < \delta < 1$, $c$ is $(\varepsilon,\delta)$-DP where $\varepsilon$ is defined by 
\[
\varepsilon = 
\begin{cases}
(\rho + \log(1-\delta))_{+} & 1 < \alpha\delta\\
\min \left(
\begin{array}{l@{}}
\left((\rho - \frac{\log \delta + \log \alpha}{\alpha-1}+ \log\left(\frac{\alpha-1}{\alpha}\right)\right)_{+},\\
\frac{1}{\alpha - 1} \log \left(\frac{\exp (\rho (\alpha - 1)) - 1}{\alpha \delta} + 1 \right)
\end{array}
\right)
& \alpha\delta \leq 1
\end{cases}
\]
\end{lemma}
This conversion law is the best one we know, but the proof is much harder than the previous two lemmas.
We formalize all these conversion laws by using the \emph{privacy regions} for DP and RDP~\cite{KairouzOV15,DBLP:journals/corr/abs-1905-09982} and reducing the conversion of divergences to the inclusion of privacy regions.
We see the details later.
\paragraph{Moment accountants in DP}
R\'{e}nyi divergence $D^\alpha(\mu_1,\mu_2)$ of order $\alpha$ gives $\alpha$-moment of random variable $\frac{p_1(x)}{p_2(x)}$ ($x \sim \mu_2$), which is privacy loss random variable when $\mu_i = c(D_i)$.
Roughly speaking, R\'{e}nyi divergence can access higher-order moments of privacy loss random variables. Using this, we obtain a tighter DP bound than just applying conversion laws (for fixed $\alpha$) or advanced composition theorems in DP.

\begin{lemma}[{\cite[Sec 3.2]{AbadiCGMMT016},\cite[Sec. IV]{DBLP:journals/corr/abs-2008-06529}}]
\label{lem:RDP:moment1}
Assume $0 < \delta < 1$, $0 < \varepsilon$, and
$K \leq  \varepsilon + 2\log(1/\delta) - 2 \sqrt{(\varepsilon +\log(1/\delta))\log(1/\delta)}$.
If $c$ is $(\alpha,\alpha K)$-RDP for any $1 < \alpha$ then $c$ is $(\varepsilon,\delta)$-DP.
\end{lemma}
We formalize this lemma using Lemma \ref{convRDPtoDP:Mironov}.
A better (and more challenging to mechanize) bound is given in \cite[Sec. IV]{DBLP:journals/corr/abs-2008-06529} using Lemma \ref{convRDPtoDP:Asoodeh}.
Tracking the proof of Lemma \ref{lem:RDP:moment1}.
It is enough to see $(\hat\alpha,\hat\alpha K)$-RDP for 
\[
\hat\alpha = 
1 + \frac{\log(1/\delta) + \sqrt{(\varepsilon +\log(1/\delta))\log(1/\delta)}}{\varepsilon}.
\]
\begin{lemma}[Explicit ver.]
\label{lem:RDP:moment2}
Assume that $\varepsilon$, $\delta$, $K$ satisfy the condition in Lemma \ref{lem:RDP:moment1}.
If $c$ is $(\hat\alpha,\hat\alpha K)$-RDP then $c$ is $(\varepsilon,\delta)$-DP.
\end{lemma}

\section{DP in Isabelle/HOL}

This work rely highly on the formalization of DP in the continuous setting~\cite{10.1145/3703595.3705875,Differential_Privacy-AFP}.
The library \isa{Differetial\_Privacy} contains not only the formalization of DP, but also important constructions for the formalization of statistical divergences.
In this work, we use the following three constructions:

\paragraph{Density functions of two measures}
In general, in theories on DP and RDP, we often need density functions.
For example, as we have seen, R\'{e}nyi divergence requires a base measure $\pi$ and \emph{density functions} $p_1,p_2$ of $\mu_1,\mu_2\in \prob (X)$ with respect to $\pi$.
They can be formulated as Radon-Nikod\'{y}m derivatives $d\mu_1/d\pi$ and $d\mu_2/d\pi$ where $\pi$ is a sigma-finite measure dominating\footnote{
Let $\mu$ and $\mu$ be (sigma-)finite measures on $X$.
We say that $\nu$ dominates $\mu$ , or $\mu$ is absolutely continuous with respect to $\nu$ (written $\mu \ll \nu$) if $\forall S \in \Sigma_X. \nu(S) = 0 \implies \mu(S) = 0$.
} both the $\mu_1$ and $\mu_2$.
We remark that the sum $\pi = \mu_1+\mu_2$ is always such a base measure.

The library \isa{Differetial\_Privacy} contains the locales for density functions of two measures, and several propreties of them. 
It supports density functions between two probability measures and finite measures, but not subprobability measures.
To formalize R\'{e}nyi divergence instead of DP, we need to overhaul this part (thus we enclose the modified copy of \isa{Differetial\_Privacy} in the supplementary material).

\paragraph{Measurable spaces of lists}
The theory \isa{List\_Space} of \isa{Differetial\_Privacy}
contains the construction of measurable spaces of finite lists and measurability proofs of list operations.
This library contains the constructor
\begin{center}
\isa{listM\ {\isacharcolon}{\kern0pt}{\isacharcolon}{\kern0pt}\ {\isachardoublequoteopen}{\isacharprime}{\kern0pt}a\ measure\ {\isasymRightarrow}\ {\isacharprime}{\kern0pt}a\ list\ measure{\isachardoublequoteclose}}
\end{center}
making a measurable space \isa{M} to the 
one \isa{listM\ M} of lists over  \isa{M}. 
For example, the space $\Real^\ast$ of lists of real numbers is denoted by \isa{listM\ borel}. 
This library contains measurability of list operations.
For example, measurability \isa{foldl} is, 
\begin{isabelle}
\isakeywordONE{lemma}\isamarkupfalse%
\ measurable{\isacharunderscore}{\kern0pt}foldl\ {\isacharbrackleft}{\kern0pt}measurable{\isacharbrackright}{\kern0pt}{\isacharcolon}{\kern0pt}\isanewline
\ \ \isakeywordTWO{assumes}\ {\isacharbrackleft}{\kern0pt}measurable{\isacharbrackright}{\kern0pt}{\isacharcolon}{\kern0pt}{\isachardoublequoteopen}{\isacharparenleft}{\kern0pt}{\isasymlambda}{\isacharparenleft}{\kern0pt}x{\isacharcomma}{\kern0pt}y{\isacharparenright}{\kern0pt}{\isachardot}{\kern0pt}\ F\ y\ x{\isacharparenright}{\kern0pt}\ {\isasymin}\ {\isacharparenleft}{\kern0pt}M\ {\isasymOtimes}\isactrlsub M\ N\ {\isasymrightarrow}\isactrlsub M\ N{\isacharparenright}{\kern0pt}{\isachardoublequoteclose}\isanewline
\ \ \isakeywordTWO{shows}\ {\isachardoublequoteopen}{\isacharparenleft}{\kern0pt}{\isasymlambda}\ {\isacharparenleft}{\kern0pt}T{\isacharcomma}{\kern0pt}xs{\isacharparenright}{\kern0pt}{\isachardot}{\kern0pt}\ foldl\ F\ T\ xs{\isacharparenright}{\kern0pt}\ {\isasymin}\ N\ {\isasymOtimes}\isactrlsub M\ {\isacharparenleft}{\kern0pt}listM\ M{\isacharparenright}{\kern0pt}\ {\isasymrightarrow}\isactrlsub M\ N{\isachardoublequoteclose}
\end{isabelle}
Furthermore, measurability of \isa{Cons}, \isa{map}, \isa{(\isacharat)}, \isa{foldr}, \isa{foldl}, \isa{fold}, \isa{take}, \isa{drop} and \isa{nth} are provided in similar ways. 

\paragraph{Formalization of DP}
The library, of course, contains formalizations of DP in the continuous setting.
Moreover, it first gives the formalization of divergence $\DP \varepsilon X (\mu,\nu)$ and its basic properties, and then formalizes DP.
\begin{isabelle}
\isacommand{definition}\isamarkupfalse%
\ DP{\isacharunderscore}{\kern0pt}divergence\ \isakeyword{where}\isanewline
\ \ {\isachardoublequoteopen}DP{\isacharunderscore}{\kern0pt}divergence\ M\ N\ {\isasymepsilon}\ {\isacharequal}{\kern0pt}\ Sup\ {\isacharbraceleft}{\kern0pt}ereal{\isacharparenleft}{\kern0pt}measure\ M\ A\ {\isacharminus}{\kern0pt}\ {\isacharparenleft}{\kern0pt}exp\ {\isasymepsilon}{\isacharparenright}{\kern0pt}\ {\isacharasterisk}{\kern0pt}\ measure\ N\ A{\isacharparenright}{\kern0pt}\ {\isacharbar}{\kern0pt}\ A{\isacharcolon}{\kern0pt}{\isacharcolon}{\kern0pt}{\isacharprime}{\kern0pt}a\ set{\isachardot}{\kern0pt}\ A\ {\isasymin}\ {\isacharparenleft}{\kern0pt}sets\ M{\isacharparenright}{\kern0pt}{\isacharbraceright}{\kern0pt}{\isachardoublequoteclose}
\end{isabelle}

\begin{isabelle}
\isacommand{definition}\isamarkupfalse%
\ DP{\isacharunderscore}{\kern0pt}inequality\ \isakeyword{where}\isanewline
\ \ {\isachardoublequoteopen}DP{\isacharunderscore}{\kern0pt}inequality\ M\ N\ {\isasymepsilon}\ {\isasymdelta}\ {\isasymequiv}\ {\isacharparenleft}{\kern0pt}{\isasymforall}\ A\ {\isasymin}\ sets\ M{\isachardot}{\kern0pt}\ measure\ M\ A\ {\isasymle}\ {\isacharparenleft}{\kern0pt}exp\ {\isasymepsilon}{\isacharparenright}{\kern0pt}\ {\isacharasterisk}{\kern0pt}\ measure\ N\ A\ {\isacharplus}{\kern0pt}\ {\isasymdelta}{\isacharparenright}{\kern0pt}{\isachardoublequoteclose}
\end{isabelle}

\begin{isabelle}
\isacommand{definition}\isamarkupfalse%
\ differential{\isacharunderscore}{\kern0pt}privacy\ \isakeyword{where}\isanewline
\ \ {\isachardoublequoteopen}differential{\isacharunderscore}{\kern0pt}privacy\ M\ adj\ {\isasymepsilon}\ {\isasymdelta}\ {\isasymequiv}\ {\isasymforall}{\isacharparenleft}{\kern0pt}d{\isadigit{1}}{\isacharcomma}{\kern0pt}d{\isadigit{2}}{\isacharparenright}{\kern0pt}{\isasymin}adj{\isachardot}{\kern0pt}\ DP{\isacharunderscore}{\kern0pt}inequality\ {\isacharparenleft}{\kern0pt}M\ d{\isadigit{1}}{\isacharparenright}{\kern0pt}\ {\isacharparenleft}{\kern0pt}M\ d{\isadigit{2}}{\isacharparenright}{\kern0pt}\ {\isasymepsilon}\ {\isasymdelta}\ {\isasymand}\ DP{\isacharunderscore}{\kern0pt}inequality\ {\isacharparenleft}{\kern0pt}M\ d{\isadigit{2}}{\isacharparenright}{\kern0pt}\ {\isacharparenleft}{\kern0pt}M\ d{\isadigit{1}}{\isacharparenright}{\kern0pt}\ {\isasymepsilon}\ {\isasymdelta}{\isachardoublequoteclose}
\end{isabelle}
Here, \isa{differential{\isacharunderscore}{\kern0pt}privacy\ M\ adj\ {\isasymepsilon}\ {\isasymdelta}} means that a mechanism \isa{M} is (\isa{\isasymepsilon},\isa{\isasymdelta})-DP with respect to the adjacency relation \isa{adj}.

In the library, formalizations of the Laplace mechanism and Report-Noisy-Max mechanism are given.
However, the Gaussian mechanism and RDP in the continuous setting are not formalized yet.
In this paper, we propose a library for the Gaussian mechanism and RDP, and give a formalization of the gradient descent mechanism with subsampling.
\section{$f$-divergence in Isabelle/HOL}
We formalize a class of divergences called \emph{$f$-divergences}~\cite{Csiszar67,1705001_2006} as a technical basis of R\'{e}nyi divergence.
For a given convex function $f \colon (0,\infty) \to \Real$, called a \emph{generator function}, we define the $\Delta^{\smash{f}}$ as follows: for any measurable space $X$ and probability measures $\mu_1,\mu_2\in \prob(X)$,
\[
\Delta^{\smash{f}}_X(\mu_1,\mu_2) = \int f\left( \frac{p_1(x)}{p_2(x)} \right) p_2(x) ~d\pi(x).
\]
Here, $p_1(x)$ and $p_2(x)$ are density functions of $\mu_1$ and $\mu_2$ with respect to a base measure $\pi$.

The divergence $\Delta^{\smash{f}}_X(\mu_1,\mu_2)$ may not be integrable (it can be $\infty$).
Technically speaking, the generator function $f$ inside the integtal is \emph{extended} by 
$f(0) = \lim_{t\to +0}f(t)$ and $0f(a/0) = a f^\ast(0)$
where $f^\ast (t) = t f(1/t)$ is the convex conjugate of $f$. 

Here, R\'{e}nyi divergence can be reformulated by 
\[
D^\alpha_X(\mu_1,\mu_2) = \frac{1}{\alpha - 1}\log \Delta^{f^\alpha}_X(\mu_1,\mu_2)\qquad (1<\alpha)
\]
where the generator function is given by $f^\alpha(t) = t^\alpha$. 

In this paper, we formalize $f$-divergences for nonnegative generator functions ($\forall t \in (0,\infty). 0 \leq f(t)$).
It is enough to formalize R\'{e}nyi divergences since $f(t) = t^\alpha$ ($1< \alpha$) is nonnegative.
It is easier to mechanize; we can skip partitioning the positive and negative parts in the integral in $\Delta^{\smash{f}}_X(\mu_1,\mu_2)$.
For probability measures, without loss of generality, we then assume that the generator function is nonnegative.
For two convex functions $f, g \colon (0,\infty) \to \Real$, if $\forall t \in (0,\infty). g(t) = f(t) + c(t -1)$ then $\Delta^{\smash{f}}_X(\mu_1,\mu_2) = \Delta^{\smash{g}}_X(\mu_1,\mu_2)$.
For a general convex $f$, we can choose $c$ such that 
$g(t) = f(t) + c(t -1)$ is nonnegative.
Therefore, if needed, we can extend basic results of $f$-divergences for nonnegative generator functions to general ones.

In this section, we formalize $f$-divergences for nonnegative generator functions as follows: we first formalize basic results for generator functions.
Then, we formalize $f$-divergences for nonnegative generator functions and give formal proofs of their basic properties and special properties of $f$-divergences, which we call continuity and approximability, used in the composability of R\'{e}nyi divergence.
\subsection{Convex real functions}
We formalize generator functions $f \colon (0,\infty) \to \Real$ and their basic results.
For example, such $f$ is \emph{piecewisely monotone}:
\begin{enumerate}
\item $f$ is monotone increasing on $(0,\infty)$,
\item $f$ is monotone decreasing on $(0,\infty)$, or
\item There is $z \in (0,\infty)$ such that $f$ is monotone decreasing on $(0,z]$ and $f$ is monotone increasing on $[z,\infty)$.
\end{enumerate}
We need it to formalize the well-definedness of the extension $f(0) = \lim_{t\to +0}f(t)$.
If $f$ were two-sided differentiable everywhere on $(0,\infty)$, the proof of piecewise monotonicity would be easy, but it is not in general.
Thus, to prove it, we use left and right derivatives instead of two-sided derivatives.

We formalize the existence of left and right derivatives of a convex function $f \colon I \to \Real$ ($I$ is a convex open subset of $\Real$), their monotonicity, and inequalities about them.
For example, $D_L f(v) = \lim_{t \to v+0} \frac{f(t)-f(v)}{t-v} \leq \frac{f(w)-f(v)}{w-v}$ ($v < w$):
\begin{isabelle}
\isakeywordONE{abbreviation}\isamarkupfalse%
\ {\isachardoublequoteopen}slope{\isacharunderscore}{\kern0pt}of\ f\ a\ b\ {\isasymequiv}\ {\isacharparenleft}{\kern0pt}f\ a\ {\isacharminus}{\kern0pt}\ f\ b{\isacharparenright}{\kern0pt}\ {\isacharslash}{\kern0pt}\ {\isacharparenleft}{\kern0pt}a\ {\isacharminus}{\kern0pt}\ b{\isacharparenright}{\kern0pt}{\isachardoublequoteclose}
\end{isabelle}
\begin{isabelle}
\isakeywordONE{lemma}\isamarkupfalse%
\ convex{\isacharunderscore}{\kern0pt}right{\isacharunderscore}{\kern0pt}deriv{\isacharunderscore}{\kern0pt}slope{\isacharunderscore}{\kern0pt}upper{\isacharcolon}{\kern0pt}\isanewline
\ \ \isakeywordTWO{fixes}\ f\ {\isacharcolon}{\kern0pt}{\isacharcolon}{\kern0pt}\ {\isachardoublequoteopen}real\ {\isasymRightarrow}\ real{\isachardoublequoteclose}\ \isakeywordTWO{assumes}\ {\isachardoublequoteopen}convex{\isacharunderscore}{\kern0pt}on\ I\ f{\isachardoublequoteclose}\ \isakeywordTWO{and}\ {\isachardoublequoteopen}v\ {\isasymin}\ I{\isachardoublequoteclose}\ \isakeywordTWO{and}\ {\isachardoublequoteopen}w\ {\isasymin}\ I{\isachardoublequoteclose}\isanewline
\ \ \ \ \isakeywordTWO{and}\ {\isachardoublequoteopen}{\isacharparenleft}{\kern0pt}f\ has{\isacharunderscore}{\kern0pt}field{\isacharunderscore}{\kern0pt}derivative\ Dv{\isacharparenright}{\kern0pt}\ {\isacharparenleft}{\kern0pt}at{\isacharunderscore}{\kern0pt}right\ v{\isacharparenright}{\kern0pt}{\isachardoublequoteclose}\ \isakeywordTWO{and}\  {\isachardoublequoteopen}v\ {\isacharless}{\kern0pt}\ w{\isachardoublequoteclose}\isanewline
\ \ \isakeywordTWO{shows}\ {\isachardoublequoteopen}Dv\ {\isasymle}\ slope{\isacharunderscore}{\kern0pt}of\ f\ w\ v{\isachardoublequoteclose}
\end{isabelle}
%
We then formalize the piecewise motonicity.
\begin{isabelle}
\isakeywordONE{theorem}\isamarkupfalse%
\ convex{\isacharunderscore}{\kern0pt}piecewise{\isacharunderscore}{\kern0pt}monotone{\isacharcolon}{\kern0pt}\isanewline
\ \ \isakeywordTWO{fixes}\ f\ {\isacharcolon}{\kern0pt}{\isacharcolon}{\kern0pt}\ {\isachardoublequoteopen}real\ {\isasymRightarrow}\ real{\isachardoublequoteclose}\ \isakeywordTWO{assumes}\ convex{\isacharunderscore}{\kern0pt}on\ I\ f{\isachardoublequoteclose}\ \isakeywordTWO{and}\ {\isachardoublequoteopen}open\ I{\isachardoublequoteclose}\isanewline
\ \ \isakeywordTWO{shows}\ {\isachardoublequoteopen}{\isacharparenleft}{\kern0pt}{\isasymexists}v\ {\isasymin}\ I{\isachardot}{\kern0pt}\ antimono{\isacharunderscore}{\kern0pt}on\ {\isacharparenleft}{\kern0pt}I\ {\isasyminter}\ {\isacharbraceleft}{\kern0pt}{\isachardot}{\kern0pt}{\isachardot}{\kern0pt}v{\isacharbraceright}{\kern0pt}{\isacharparenright}{\kern0pt}\ f\ {\isasymand}\ mono{\isacharunderscore}{\kern0pt}on\ \ {\isacharparenleft}{\kern0pt}I\ {\isasyminter}\ {\isacharbraceleft}{\kern0pt}v{\isachardot}{\kern0pt}{\isachardot}{\kern0pt}{\isacharbraceright}{\kern0pt}{\isacharparenright}{\kern0pt}\ f\ {\isacharparenright}{\kern0pt}\ {\isasymor}\ {\isacharparenleft}{\kern0pt}mono{\isacharunderscore}{\kern0pt}on\ I\ f{\isacharparenright}{\kern0pt}\ {\isasymor}\ {\isacharparenleft}{\kern0pt}antimono{\isacharunderscore}{\kern0pt}on\ I\ f{\isacharparenright}{\kern0pt}{\isachardoublequoteclose}
\end{isabelle}
From these lemmas and the Froda theorem (formalized as \isa{mono\_on\_ctble\_discont\_open} in the standard library), every convex real function is differentiable almost everywhere.
\begin{isabelle}
\isakeywordONE{lemma}\isamarkupfalse%
\ convex{\isacharunderscore}{\kern0pt}differentialble{\isacharunderscore}{\kern0pt}AE{\isacharcolon}{\kern0pt}\isanewline
\ \ \isakeywordTWO{fixes}\ f\ {\isacharcolon}{\kern0pt}{\isacharcolon}{\kern0pt}\ {\isachardoublequoteopen}real\ {\isasymRightarrow}\ real{\isachardoublequoteclose}\ \isakeywordTWO{assumes}\ cvxf{\isacharcolon}{\kern0pt}{\isachardoublequoteopen}convex{\isacharunderscore}{\kern0pt}on\ I\ f{\isachardoublequoteclose}\ \isakeywordTWO{and}\  {\isachardoublequoteopen}open\ I{\isachardoublequoteclose}\isanewline
\ \ \isakeywordTWO{shows}\ {\isachardoublequoteopen}AE\ x\ {\isasymin}\ I\ in\ lborel{\isachardot}{\kern0pt}\ f\ differentiable\ {\isacharparenleft}{\kern0pt}at\ x{\isacharparenright}{\kern0pt}{\isachardoublequoteclose}
\end{isabelle}

\paragraph{Countable version of Jensen inequality}
We also prove a countable version of Jensen inequality:
The finite version ($C$: finite) is already formalized in isabelle/HOL's standard library.
We want to support countable infinite $C$. 
\begin{isabelle}
\isakeywordONE{lemma}\isamarkupfalse%
\ convex{\isacharunderscore}{\kern0pt}on{\isacharunderscore}{\kern0pt}infsum{\isacharunderscore}{\kern0pt}real{\isacharcolon}{\kern0pt}\isanewline
\ \ \ \isakeywordTWO{fixes}\ a\ {\isacharcolon}{\kern0pt}{\isacharcolon}{\kern0pt}\ {\isachardoublequoteopen}{\isacharprime}{\kern0pt}a\ {\isasymRightarrow}\ real{\isachardoublequoteclose}\ \isakeywordTWO{and}\ y\ {\isacharcolon}{\kern0pt}{\isacharcolon}{\kern0pt}\ {\isachardoublequoteopen}{\isacharprime}{\kern0pt}a\ {\isasymRightarrow}\ real{\isachardoublequoteclose}\ \isakeywordTWO{and}\ f\ {\isacharcolon}{\kern0pt}{\isacharcolon}{\kern0pt}\ {\isachardoublequoteopen}real\ {\isasymRightarrow}\ real{\isachardoublequoteclose}\isanewline
\ \ \isakeywordTWO{assumes}\ {\isachardoublequoteopen}convex{\isacharunderscore}{\kern0pt}on\ I\ f{\isachardoublequoteclose}\ \isakeywordTWO{and}\ {\isachardoublequoteopen}open\ I{\isachardoublequoteclose}\ \isakeywordTWO{and}\ {\isachardoublequoteopen}{\isacharparenleft}{\kern0pt}a\ has{\isacharunderscore}{\kern0pt}sum\ {\isadigit{1}}{\isacharparenright}{\kern0pt}\ C{\isachardoublequoteclose}\isanewline
\ \ \ \ \isakeywordTWO{and}\ {\isachardoublequoteopen}{\isacharparenleft}{\kern0pt}{\isasymlambda}i{\isachardot}{\kern0pt}\ a\ i\ {\isacharasterisk}{\kern0pt}\ y\ i{\isacharparenright}{\kern0pt}\ summable{\isacharunderscore}{\kern0pt}on\ C{\isachardoublequoteclose}\ {\isachardoublequoteopen}{\isacharparenleft}{\kern0pt}{\isasymSum}\isactrlsub {\isasyminfinity}i{\isasymin}C{\isachardot}{\kern0pt}\ a\ i\ {\isacharasterisk}{\kern0pt}\ y\ i{\isacharparenright}{\kern0pt}\ {\isasymin}\ I{\isachardoublequoteclose}\isanewline
\ \ \ \ \isakeywordTWO{and}\ {\isachardoublequoteopen}{\isacharparenleft}{\kern0pt}{\isasymlambda}i{\isachardot}{\kern0pt}\ a\ i\ {\isacharasterisk}{\kern0pt}\ f\ {\isacharparenleft}{\kern0pt}y\ i{\isacharparenright}{\kern0pt}{\isacharparenright}{\kern0pt}\ summable{\isacharunderscore}{\kern0pt}on\ C{\isachardoublequoteclose}\ \ \isanewline
\ \ \ \ \isakeywordTWO{and}\ {\isachardoublequoteopen}{\isasymforall}i{\isasymin}C{\isachardot}{\kern0pt}\ a\ i\ {\isasymge}\ {\isadigit{0}}{\isachardoublequoteclose}\ \isakeywordTWO{and}\ {\isachardoublequoteopen}{\isasymforall}i{\isasymin}C{\isachardot}{\kern0pt}\ y\ i\ {\isasymin}\ I{\isachardoublequoteclose}\isanewline
\ \ \isakeywordTWO{shows}\ {\isachardoublequoteopen}f\ {\isacharparenleft}{\kern0pt}{\isasymSum}\isactrlsub {\isasyminfinity}i{\isasymin}C{\isachardot}{\kern0pt}\ a\ i\ {\isacharasterisk}{\kern0pt}\ y\ i{\isacharparenright}{\kern0pt}\ {\isasymle}\ {\isacharparenleft}{\kern0pt}{\isasymSum}\isactrlsub {\isasyminfinity}i{\isasymin}C{\isachardot}{\kern0pt}\ a\ i\ {\isacharasterisk}{\kern0pt}f\ {\isacharparenleft}{\kern0pt}y\ i{\isacharparenright}{\kern0pt}{\isacharparenright}{\kern0pt}{\isachardoublequoteclose}
\end{isabelle}
Here, we choose \isa{suminf}(\isa{{\isasymSum}\isactrlsub {\isasyminfinity}}) for countable sums instead of \isa{infsum}, 
because \isa{suminf} supports any type of sum $\sum_{i \in C} a_i$ though it needs the convergence (written \isa{summable}), while \isa{infsum} supports only sums indexed by natural numbers.
%

\subsection{Basics of generator function and extension}
We formalize a generator function for $f$-divergences:
\begin{isabelle}
\isakeywordONE{definition}\isamarkupfalse%
\ generator{\isacharunderscore}{\kern0pt}F\ {\isacharcolon}{\kern0pt}{\isacharcolon}{\kern0pt}\ {\isachardoublequoteopen}{\isacharparenleft}{\kern0pt}real\ {\isasymRightarrow}\ real{\isacharparenright}{\kern0pt}\ {\isasymRightarrow}\ bool{\isachardoublequoteclose}\ \isakeywordTWO{where}\isanewline
\ \ {\isachardoublequoteopen}generator{\isacharunderscore}{\kern0pt}F\ f\ {\isasymlongleftrightarrow}\ {\isacharparenleft}{\kern0pt}convex{\isacharunderscore}{\kern0pt}on\ {\isacharbraceleft}{\kern0pt}{\isadigit{0}}{\isacharless}{\kern0pt}{\isachardot}{\kern0pt}{\isachardot}{\kern0pt}{\isacharbraceright}{\kern0pt}\ f{\isacharparenright}{\kern0pt}{\isachardoublequoteclose}
\end{isabelle}
Thanks to the piecewise monotonicity, the extension of the generator function by $f(0) = \lim_{t\to +0}f(t)$ is well-defined.
\begin{isabelle}
\isakeywordONE{lemma}\isamarkupfalse%
\ gf{\isacharunderscore}{\kern0pt}converge{\isacharunderscore}{\kern0pt}to{\isacharunderscore}{\kern0pt}{\isadigit{0}}{\isacharcolon}{\kern0pt}\isanewline
\ \ \isakeywordTWO{assumes}\ {\isachardoublequoteopen}generator{\isacharunderscore}{\kern0pt}F\ f{\isachardoublequoteclose}\isanewline
\ \ \isakeywordTWO{shows}\ {\isachardoublequoteopen}{\isasymexists}\ f{\isadigit{0}}{\isacharcolon}{\kern0pt}{\isacharcolon}{\kern0pt}\ ereal{\isachardot}{\kern0pt}\ {\isacharparenleft}{\kern0pt}{\isacharparenleft}{\kern0pt}ereal\ o\ f{\isacharparenright}{\kern0pt}\ {\isasymlonglongrightarrow}\ f{\isadigit{0}}{\isacharparenright}{\kern0pt}\ {\isacharparenleft}{\kern0pt}at{\isacharunderscore}{\kern0pt}right\ {\isadigit{0}}{\isacharparenright}{\kern0pt}\ {\isasymand}\ {\isacharminus}{\kern0pt}{\isasyminfinity}\ {\isacharless}{\kern0pt}\ f{\isadigit{0}}{\isachardoublequoteclose}
\end{isabelle}
We now define the extension $\bar{f} \colon [0,\infty) \to [-\infty,\infty]$ of $f$ by $\bar{f}(t) = f (t)$ for $0 < t$ and $\bar{f}(0) = \lim_{t\to +0}f(t)$, and formalize it as follows:
\begin{isabelle}
\isakeywordONE{definition}\isamarkupfalse%
\ extend{\isacharunderscore}{\kern0pt}F\ {\isacharcolon}{\kern0pt}{\isacharcolon}{\kern0pt}\ {\isachardoublequoteopen}{\isacharparenleft}{\kern0pt}real\ {\isasymRightarrow}\ real{\isacharparenright}{\kern0pt}\ {\isasymRightarrow}\ {\isacharparenleft}{\kern0pt}real\ {\isasymRightarrow}\ ereal{\isacharparenright}{\kern0pt}{\isachardoublequoteclose}\ \isakeywordTWO{where}\isanewline
\ \ {\isachardoublequoteopen}extend{\isacharunderscore}{\kern0pt}F\ f\ x\ {\isacharequal}{\kern0pt}\ {\isacharparenleft}{\kern0pt}if\ x\ {\isacharless}{\kern0pt}\ {\isadigit{0}}\ then\ {\isadigit{0}}\ else\ if\ x\ {\isacharequal}{\kern0pt}\ {\isadigit{0}}\ then\ SOME\ f{\isadigit{0}}{\isachardot}{\kern0pt}\ {\isacharparenleft}{\kern0pt}{\isacharparenleft}{\kern0pt}ereal\ o\ f{\isacharparenright}{\kern0pt}\ {\isasymlonglongrightarrow}\ f{\isadigit{0}}{\isacharparenright}{\kern0pt}\ {\isacharparenleft}{\kern0pt}at{\isacharunderscore}{\kern0pt}right\ {\isadigit{0}}{\isacharparenright}{\kern0pt}\ else\ ereal\ {\isacharparenleft}{\kern0pt}f\ x{\isacharparenright}{\kern0pt}{\isacharparenright}{\kern0pt}{\isachardoublequoteclose}
\end{isabelle}
The definition of \isa{extend\_F} requires the existence of a limit, which we have already proved.
For inputs of negative values (outside of $[0,\infty)$), we set a dummy value of $0$. 

Next we formalize the extension $0f(a/0) = a  f^\ast(0)$ ($a \in [0,\infty)$).
We give the convex conjugate $f^\ast (t) = tf(1/t)$.
\begin{isabelle}
\isakeywordONE{definition}\isamarkupfalse%
\ conj{\isacharunderscore}{\kern0pt}F\ {\isacharcolon}{\kern0pt}{\isacharcolon}{\kern0pt}\ {\isachardoublequoteopen}{\isacharparenleft}{\kern0pt}real\ {\isasymRightarrow}\ real{\isacharparenright}{\kern0pt}\ {\isasymRightarrow}\ {\isacharparenleft}{\kern0pt}real\ {\isasymRightarrow}\ real{\isacharparenright}{\kern0pt}{\isachardoublequoteclose}\ \isakeywordTWO{where}\isanewline
\ \ {\isachardoublequoteopen}conj{\isacharunderscore}{\kern0pt}F\ f\ {\isacharequal}{\kern0pt}\ {\isacharparenleft}{\kern0pt}{\isasymlambda}t{\isachardot}{\kern0pt}\ t\ {\isacharasterisk}{\kern0pt}\ f\ {\isacharparenleft}{\kern0pt}{\isadigit{1}}\ {\isacharslash}{\kern0pt}\ t{\isacharparenright}{\kern0pt}{\isacharparenright}{\kern0pt}{\isachardoublequoteclose}
\end{isabelle}
We then introduce a new function $F_f \colon [0,\infty) \to [0,\infty) \to [-\infty,\infty]$ with two arguments for $a$ and $b$ to formalize the term $bf(a/b)$ including the case of $b = 0$.
Isabelle HOL's standard library defines $a/0 = 0$ to totalize the division operator $(/)$.
It is normal in many proof assistants, but not suitable for formalizing $0f(a/0)$.
In the definition of $f$-divergence, we use the extension $F_f$ instead of the original $f$.

We define $F_f\ a\ b = b \bar{f}(a/b)$ ($0 < b$) and $F_f\ a\ 0 =  a  f^\ast(0)$, and formalize it is as follows: 
\begin{isabelle}
\isakeywordONE{definition}\isamarkupfalse%
\ extend{\isacharunderscore}{\kern0pt}frac{\isacharunderscore}{\kern0pt}F\ {\isacharcolon}{\kern0pt}{\isacharcolon}{\kern0pt}\ {\isachardoublequoteopen}{\isacharparenleft}{\kern0pt}real\ {\isasymRightarrow}\ real{\isacharparenright}{\kern0pt}\ {\isasymRightarrow}\ {\isacharparenleft}{\kern0pt}real\ {\isasymRightarrow}\ real\ {\isasymRightarrow}\ ereal{\isacharparenright}{\kern0pt}{\isachardoublequoteclose}\ \isakeywordTWO{where}
\ {\isachardoublequoteopen}extend{\isacharunderscore}{\kern0pt}frac{\isacharunderscore}{\kern0pt}F\ f\ a\ b\ {\isacharequal}{\kern0pt}\ {\isacharparenleft}{\kern0pt}if\ a\ {\isacharless}{\kern0pt}\ {\isadigit{0}}\ then\ undefined\isanewline
\ else\ if\ b\ {\isacharless}{\kern0pt}\ {\isadigit{0}}\ then\ undefined\isanewline
\ else\ if\ a\ {\isacharequal}{\kern0pt}\ {\isadigit{0}}\ then\ b\ {\isacharasterisk}{\kern0pt}\ {\isacharparenleft}{\kern0pt}extend{\isacharunderscore}{\kern0pt}F\ f\ {\isadigit{0}}{\isacharparenright}{\kern0pt}\isanewline
\ else\ if\ b\ {\isacharequal}{\kern0pt}\ {\isadigit{0}}\ then\ a\ {\isacharasterisk}{\kern0pt}\ {\isacharparenleft}{\kern0pt}extend{\isacharunderscore}{\kern0pt}F\ {\isacharparenleft}{\kern0pt}conj{\isacharunderscore}{\kern0pt}F\ f{\isacharparenright}{\kern0pt}\ {\isadigit{0}}{\isacharparenright}{\kern0pt}
\ else\ b\ {\isacharasterisk}{\kern0pt}\ f\ {\isacharparenleft}{\kern0pt}a{\isacharslash}{\kern0pt}b{\isacharparenright}{\kern0pt}{\isacharparenright}{\kern0pt}{\isachardoublequoteclose}
\end{isabelle}
For $a, b \notin [0,\infty)$ \isa{extend{\isacharunderscore}{\kern0pt}frac{\isacharunderscore}{\kern0pt}F\ a\ b} returns \isa{undefined}.

\paragraph{Extension of countable Jensen inequality}
We formalize a variant of Jensen inequality
\[
{\textstyle \left({\sum_{i \in C} a_i}\right)} f\left(\frac{\sum_{i \in C} x_i}{\sum_{i \in C} a_i}\right) \leq 
\sum_{i \in C} a_i  f\left(\frac{x_i}{a_i} \right) \quad (\forall i \in C. 0 \leq a_i, x_i)
\]
using the extension \isa{extend{\isacharunderscore}{\kern0pt}frac{\isacharunderscore}{\kern0pt}F}.
\begin{isabelle}
\isakeywordONE{lemma}\isamarkupfalse%
\ extend{\isacharunderscore}{\kern0pt}frac{\isacharunderscore}{\kern0pt}F{\isacharunderscore}{\kern0pt}convex{\isacharunderscore}{\kern0pt}on{\isacharunderscore}{\kern0pt}frac{\isacharunderscore}{\kern0pt}infsum{\isacharcolon}{\kern0pt}\isanewline
\ \ \isakeywordTWO{fixes}\ a\ {\isacharcolon}{\kern0pt}{\isacharcolon}{\kern0pt}\ {\isachardoublequoteopen}{\isacharprime}{\kern0pt}a\ {\isasymRightarrow}\ real{\isachardoublequoteclose}\ \isakeywordTWO{and}\ y\ {\isacharcolon}{\kern0pt}{\isacharcolon}{\kern0pt}\ {\isachardoublequoteopen}{\isacharprime}{\kern0pt}a\ {\isasymRightarrow}\ real{\isachardoublequoteclose}\isanewline
\ \ \isakeywordTWO{assumes}\ gf{\isacharcolon}{\kern0pt}\ {\isachardoublequoteopen}generator{\isacharunderscore}{\kern0pt}F\ f{\isachardoublequoteclose}\isanewline
\ \ \ \ \isakeywordTWO{and}\ suma{\isacharcolon}{\kern0pt}\ {\isachardoublequoteopen}a\ summable{\isacharunderscore}{\kern0pt}on\ C{\isachardoublequoteclose}\ \isakeywordTWO{and}\ sumy{\isacharcolon}{\kern0pt}\ {\isachardoublequoteopen}y\ summable{\isacharunderscore}{\kern0pt}on\ C{\isachardoublequoteclose}\isanewline
\ \ \ \ \isakeywordTWO{and}\ {\isachardoublequoteopen}{\isasymforall}i{\isasymin}C{\isachardot}{\kern0pt}\ a\ i\ {\isasymge}\ {\isadigit{0}}{\isachardoublequoteclose}\
 \isakeywordTWO{and}\ {\isachardoublequoteopen}{\isasymforall}i{\isasymin}C{\isachardot}{\kern0pt}\ y\ i\ {\isasymge}\ {\isadigit{0}}{\isachardoublequoteclose}\isanewline
\ \ \isakeywordTWO{shows}\ {\isachardoublequoteopen}extend{\isacharunderscore}{\kern0pt}frac{\isacharunderscore}{\kern0pt}F\ f\ {\isacharparenleft}{\kern0pt}infsum\ y\ C{\isacharparenright}{\kern0pt}\ {\isacharparenleft}{\kern0pt}infsum\ a\ C{\isacharparenright}{\kern0pt}\ {\isasymle}\ {\isacharparenleft}{\kern0pt}{\isasymSum}\isactrlsub {\isasyminfinity}i{\isasymin}C{\isachardot}{\kern0pt}\ extend{\isacharunderscore}{\kern0pt}frac{\isacharunderscore}{\kern0pt}F\ f\ {\isacharparenleft}{\kern0pt}y\ i{\isacharparenright}{\kern0pt}\ {\isacharparenleft}{\kern0pt}a\ i{\isacharparenright}{\kern0pt}{\isacharparenright}{\kern0pt}{\isachardoublequoteclose}
\end{isabelle}
We use this lemma to formalize approximations of $f$-divergences, which are important for formalizing R\'{e}nyi divergence.

The formal proof is done by partitioning $C$ into $\{i \in C ~|~ a_i = 0\}$ and $\{i \in C ~|~ 0 < a_i \}$, and 
extending the convexity of $f\colon(0,\infty) \to \Real$ to ones of $\bar{f}$ and $F_f$.

\subsection{Basics of $f$-divergence}
We now formalize the $f$-divergence $\Delta^{\smash{f}}$ for a nonnegative generator function $f$. We use the extension $F_f$ instead of $f$.
\[
\Delta^{\smash{f}}_X(\mu_1,\mu_2) = \int F_f \ {p_1(x)}\ {p_2(x)}~d\pi(x).
\]
\begin{isabelle}
\isakeywordONE{definition}\isamarkupfalse%
\ f{\isacharunderscore}{\kern0pt}div{\isacharunderscore}{\kern0pt}pos
\ \isakeywordTWO{where}\isanewline
\ \ {\isachardoublequoteopen}f{\isacharunderscore}{\kern0pt}div{\isacharunderscore}{\kern0pt}pos\ f\ L\ M\ N\ {\isacharequal}{\kern0pt}\ {\isasymintegral}\isactrlsup {\isacharplus}{\kern0pt}x{\isachardot}{\kern0pt}\ e{\isadigit{2}}ennreal\ {\isacharparenleft}{\kern0pt}{\isacharparenleft}{\kern0pt}extend{\isacharunderscore}{\kern0pt}frac{\isacharunderscore}{\kern0pt}F\ f{\isacharparenright}{\kern0pt}\ {\isacharparenleft}{\kern0pt}real{\isacharunderscore}{\kern0pt}RN{\isacharunderscore}{\kern0pt}deriv\ L\ M\ x{\isacharparenright}{\kern0pt}\ {\isacharparenleft}{\kern0pt}real{\isacharunderscore}{\kern0pt}RN{\isacharunderscore}{\kern0pt}deriv\ L\ N\ x{\isacharparenright}{\kern0pt}{\isacharparenright}{\kern0pt}\ {\isasympartial}L{\isachardoublequoteclose}
\end{isabelle}
Here, \isa{L}, \isa{M}, \isa{N} are $\pi$, $\mu_1$ and $\mu_2$ respectively.
We use \isa{extend{\isacharunderscore}{\kern0pt}frac{\isacharunderscore}{\kern0pt}F} to formalize $F_f$.
\isa{real{\isacharunderscore}{\kern0pt}RN{\isacharunderscore}{\kern0pt}deriv}
is construction of density functions (real-valued Radon-Nikod\'{y}m derivatives) of $M$ and $N$.

We also formalize the $f$-divergences \emph{over} a subset $A \in \Sigma_X$.  
\[
\Delta^{\smash{f}}_X|_A(\mu_1,\mu_2) = \int_A  F_f \ {p_1(x)}\ {p_2(x)}~d\pi(x).
\]
\begin{isabelle}
\isakeywordONE{definition}\isamarkupfalse%
\ f{\isacharunderscore}{\kern0pt}div{\isacharunderscore}{\kern0pt}pos{\isacharunderscore}{\kern0pt}sub
\ \isakeywordTWO{where}\isanewline
\ \ {\isachardoublequoteopen}f{\isacharunderscore}{\kern0pt}div{\isacharunderscore}{\kern0pt}pos{\isacharunderscore}{\kern0pt}sub\ f\ L\ M\ N\ A\ {\isacharequal}{\kern0pt}\ {\isacharparenleft}{\kern0pt}{\isasymintegral}\isactrlsup {\isacharplus}{\kern0pt}x{\isasymin}A{\isachardot}{\kern0pt}\ e{\isadigit{2}}ennreal\ {\isacharparenleft}{\kern0pt}{\isacharparenleft}{\kern0pt}extend{\isacharunderscore}{\kern0pt}frac{\isacharunderscore}{\kern0pt}F\ f{\isacharparenright}{\kern0pt}\ {\isacharparenleft}{\kern0pt}real{\isacharunderscore}{\kern0pt}RN{\isacharunderscore}{\kern0pt}deriv\ L\ M\ x{\isacharparenright}{\kern0pt}\ {\isacharparenleft}{\kern0pt}real{\isacharunderscore}{\kern0pt}RN{\isacharunderscore}{\kern0pt}deriv\ L\ N\ x{\isacharparenright}{\kern0pt}{\isacharparenright}{\kern0pt}\ {\isasympartial}L{\isacharparenright}{\kern0pt}{\isachardoublequoteclose}
\end{isabelle}
\begin{isabelle}
\isakeywordONE{lemma}\isamarkupfalse%
\ f{\isacharunderscore}{\kern0pt}div{\isacharunderscore}{\kern0pt}pos{\isacharunderscore}{\kern0pt}sub{\isacharunderscore}{\kern0pt}sp{\isacharcolon}{\kern0pt}\isanewline
\ \ \isakeywordTWO{shows}\ {\isachardoublequoteopen}f{\isacharunderscore}{\kern0pt}div{\isacharunderscore}{\kern0pt}pos\ f\ L\ M\ N\ {\isacharequal}{\kern0pt}\ f{\isacharunderscore}{\kern0pt}div{\isacharunderscore}{\kern0pt}pos{\isacharunderscore}{\kern0pt}sub\ f\ L\ M\ N\ {\isacharparenleft}{\kern0pt}space\ L{\isacharparenright}{\kern0pt}{\isachardoublequoteclose}
\end{isabelle}
In the formalization of basic properties of $\Delta^{\smash{f}}_X$, we often formalize ones of $\Delta^{\smash{f}}_X|_A$.

The nonnegativity $0 \leq \Delta^{\smash{f}}_X((\mu_1,\mu_2)$ is obvious:
we consider nonnegative $f$, and density functions are nonnegative.  
We also formalizet the reversal law $\Delta^{\smash{f^\ast}}_X((\mu_1,\mu_2) = \Delta^{\smash{f}}_X((\mu_2,\mu_1)$.

\paragraph{Well-defiinedness}
We formalize the well-definedness with respect to base measures.
The formal proof is based on the following facts.
If $\mu_1,\mu_2 \ll \pi$ then $\mu_1 + \mu_2 \ll \pi$.
Hence $\frac{d\mu_k}{d\pi} = \frac{d\mu_k}{d(\mu_1+\mu_2)}\frac{d(\mu_1+\mu_2)}{d\pi}$.
This implies that $f$-divergence with base measure $\pi$ is equal to the one with base measure $\mu_1 + \mu_2$.
\begin{isabelle}
\isakeywordONE{lemma}\isamarkupfalse%
\ {\isacharparenleft}{\kern0pt}\isakeywordTWO{in}\ densities{\isacharunderscore}{\kern0pt}two{\isacharunderscore}{\kern0pt}measures{\isacharunderscore}{\kern0pt}finite{\isacharparenright}{\kern0pt}\ f{\isacharunderscore}{\kern0pt}div{\isacharunderscore}{\kern0pt}pos{\isacharunderscore}{\kern0pt}well{\isacharunderscore}{\kern0pt}defined{\isacharprime}{\kern0pt}{\isacharcolon}{\kern0pt}\isanewline
\ \ \isakeywordTWO{assumes}\ gf{\isacharcolon}{\kern0pt}\ {\isachardoublequoteopen}generator{\isacharunderscore}{\kern0pt}F\ f{\isachardoublequoteclose}\ \isakeywordTWO{and}\ posf{\isacharcolon}{\kern0pt}\ {\isachardoublequoteopen}{\isasymforall}x\ {\isachargreater}{\kern0pt}\ {\isadigit{0}}{\isachardot}{\kern0pt}\ f\ x\ {\isasymge}\ {\isadigit{0}}{\isachardoublequoteclose}\isanewline
\ \ \isakeywordTWO{shows}\ {\isachardoublequoteopen}f{\isacharunderscore}{\kern0pt}div{\isacharunderscore}{\kern0pt}pos\ f\ {\isacharparenleft}{\kern0pt}sum{\isacharunderscore}{\kern0pt}measure\ M\ N{\isacharparenright}{\kern0pt}\ M\ N\ {\isacharequal}{\kern0pt}\ f{\isacharunderscore}{\kern0pt}div{\isacharunderscore}{\kern0pt}pos\ f\ L\ M\ N{\isachardoublequoteclose}
\end{isabelle}
Here, the locale \isa{densities{\isacharunderscore}{\kern0pt}two{\isacharunderscore}{\kern0pt}measures{\isacharunderscore}{\kern0pt}finite} is to give a situation to construct density functions of \isa{M} and \isa{N} with respect to \isa{L}.
\isa{sum{\isacharunderscore}{\kern0pt}measure\ M\ N} is the sum of \isa{M} and \isa{N}.

\paragraph{$f$-divergence between the same distributions}
We have $\Delta^{\smash{f}}_X(\mu,\mu) = f(1)$ for any $\mu \in \prob(X)$.
We formalize $\Delta^{\smash{f}}_X|_A(\mu,\mu) = f(1) \mu(A)$ for finite mesure $\mu$ on $X$ and $A \in \Sigma_X$, 
where apply the fact $F_f\ z\ z= (\mathrm{if}~ z = 0 ~\mathrm{then}~0~\mathrm{else}~f(1))$.

Then we formalize $D^\alpha_X(\mu,\mu) = 0$ by applying $f^\alpha(1) = 1$.

\begin{isabelle}
\isakeywordONE{lemma}\isamarkupfalse%
\ f{\isacharunderscore}{\kern0pt}div{\isacharunderscore}{\kern0pt}pos{\isacharunderscore}{\kern0pt}sub{\isacharunderscore}{\kern0pt}refl{\isacharcolon}{\kern0pt}\isanewline
\ \ \isakeywordTWO{fixes}\ a{\isacharcolon}{\kern0pt}{\isacharcolon}{\kern0pt}real\ \isakeywordTWO{assumes}\ f{\isadigit{1}}{\isadigit{0}}{\isacharcolon}{\kern0pt}\ {\isachardoublequoteopen}f\ {\isadigit{1}}\ {\isacharequal}{\kern0pt}\ a{\isachardoublequoteclose}\ \isakeywordTWO{and}\ {\isacharbrackleft}{\kern0pt}measurable{\isacharbrackright}{\kern0pt}{\isacharcolon}{\kern0pt}\ {\isachardoublequoteopen}A\ {\isasymin}\ sets\ L{\isachardoublequoteclose}\ \isanewline
\ \ \ \ \isakeywordTWO{and}\ {\isachardoublequoteopen}sigma{\isacharunderscore}{\kern0pt}finite{\isacharunderscore}{\kern0pt}measure\ L{\isachardoublequoteclose}\ {\isachardoublequoteopen}absolutely{\isacharunderscore}{\kern0pt}continuous\ L\ M{\isachardoublequoteclose}\ {\isachardoublequoteopen}finite{\isacharunderscore}{\kern0pt}measure\ M{\isachardoublequoteclose}\ {\isachardoublequoteopen}sets\ M\ {\isacharequal}{\kern0pt}\ sets\ L{\isachardoublequoteclose}\isanewline
\ \ \isakeywordTWO{shows}\ {\isachardoublequoteopen}f{\isacharunderscore}{\kern0pt}div{\isacharunderscore}{\kern0pt}pos{\isacharunderscore}{\kern0pt}sub\ f\ L\ M\ M\ A\ {\isacharequal}{\kern0pt}\ a\ {\isacharasterisk}{\kern0pt}\ emeasure\ M\ A{\isachardoublequoteclose}
\end{isabelle}
\paragraph{Joint convexity}
We formalize the joint convexity of $\Delta^{\smash{f}}$.
\begin{align*}
&\Delta^{\smash{f}}_X((1-q)\mu_1+q\nu_1,(1-q)\mu_2+q\nu_2)\\
&\qquad \leq (1-q) \Delta^{\smash{f}}_X(\mu_1,\mu_2) + q \Delta^{\smash{f}}_X(\nu_1,\nu_2) \qquad (0 \leq q \leq 1).
\end{align*}
\begin{isabelle}
\isakeywordONE{abbreviation}\isamarkupfalse%
\ {\isachardoublequoteopen}conv{\isacharunderscore}{\kern0pt}measure\ a\ M\ N\ {\isasymequiv}\ sum{\isacharunderscore}{\kern0pt}measure\ {\isacharparenleft}{\kern0pt}scale{\isacharunderscore}{\kern0pt}measure\ {\isacharparenleft}{\kern0pt}ennreal\ a{\isacharparenright}{\kern0pt}\ M{\isacharparenright}{\kern0pt}\ {\isacharparenleft}{\kern0pt}scale{\isacharunderscore}{\kern0pt}measure\ {\isacharparenleft}{\kern0pt}ennreal\ {\isacharparenleft}{\kern0pt}{\isadigit{1}}\ {\isacharminus}{\kern0pt}\ a{\isacharparenright}{\kern0pt}{\isacharparenright}{\kern0pt}\ N{\isacharparenright}{\kern0pt}{\isachardoublequoteclose}
\end{isabelle}
\begin{isabelle}
\isakeywordONE{lemma}\isamarkupfalse%
\ f{\isacharunderscore}{\kern0pt}div{\isacharunderscore}{\kern0pt}pos{\isacharunderscore}{\kern0pt}joint{\isacharunderscore}{\kern0pt}convex{\isacharcolon}{\kern0pt}\isanewline
\isakeywordTWO{fixes}\ a{\isacharcolon}{\kern0pt}{\isacharcolon}{\kern0pt}real\isanewline
\ \ \isakeywordTWO{assumes}\ gf{\isacharcolon}{\kern0pt}\ {\isachardoublequoteopen}generator{\isacharunderscore}{\kern0pt}F\ f{\isachardoublequoteclose}\
 \isakeywordTWO{and}\ posf{\isacharcolon}{\kern0pt}\ {\isachardoublequoteopen}{\isasymforall}x\ {\isachargreater}{\kern0pt}\ {\isadigit{0}}{\isachardot}{\kern0pt}\ {\isadigit{0}}\ {\isasymle}\ f\ x{\isachardoublequoteclose}\isanewline
\ \ \ \ \isakeywordTWO{and}\ cont{\isadigit{1}}{\isacharcolon}{\kern0pt}\ {\isachardoublequoteopen}densities{\isacharunderscore}{\kern0pt}two{\isacharunderscore}{\kern0pt}measures{\isacharunderscore}{\kern0pt}finite\ L\ M{\isadigit{1}}\ N{\isadigit{1}}{\isachardoublequoteclose}\isanewline
\ \ \ \ \isakeywordTWO{and}\ cont{\isadigit{2}}{\isacharcolon}{\kern0pt}\ {\isachardoublequoteopen}densities{\isacharunderscore}{\kern0pt}two{\isacharunderscore}{\kern0pt}measures{\isacharunderscore}{\kern0pt}finite\ L\ M{\isadigit{2}}\ N{\isadigit{2}}{\isachardoublequoteclose}\isanewline
\ \ \ \ \isakeywordTWO{and}\ a{\isacharcolon}{\kern0pt}\ {\isachardoublequoteopen}{\isadigit{0}}\ {\isasymle}\ a{\isachardoublequoteclose}\ {\isachardoublequoteopen}a\ {\isasymle}\ {\isadigit{1}}{\isachardoublequoteclose}\isanewline
\ \ \isakeywordTWO{shows}\ {\isachardoublequoteopen}f{\isacharunderscore}{\kern0pt}div{\isacharunderscore}{\kern0pt}pos\ f\ L\ {\isacharparenleft}{\kern0pt}conv{\isacharunderscore}{\kern0pt}measure\ a\ M{\isadigit{1}}\ M{\isadigit{2}}{\isacharparenright}{\kern0pt}\ {\isacharparenleft}{\kern0pt}conv{\isacharunderscore}{\kern0pt}measure\ a\ N{\isadigit{1}}\ N{\isadigit{2}}{\isacharparenright}{\kern0pt}\ {\isasymle}\ a\ {\isacharasterisk}{\kern0pt}\ f{\isacharunderscore}{\kern0pt}div{\isacharunderscore}{\kern0pt}pos\ f\ L\ \ M{\isadigit{1}}\ N{\isadigit{1}}\ {\isacharplus}{\kern0pt}\ {\isacharparenleft}{\kern0pt}{\isadigit{1}}{\isacharminus}{\kern0pt}a{\isacharparenright}{\kern0pt}\ {\isacharasterisk}{\kern0pt}\ f{\isacharunderscore}{\kern0pt}div{\isacharunderscore}{\kern0pt}pos\ f\ L\ M{\isadigit{2}}\ N{\isadigit{2}}{\isachardoublequoteclose}
\end{isabelle}

\subsection{Approximation by measurable partitions}
We show two kinds of approximations of $f$-divergences.
First, $f$-divergences (over $A \in \Sigma_X$) can be approximated using finite measurable partitions.
In \cite{DBLP:conf/lics/SatoBGHK19}, this is called continuity.
\begin{align*}
\Delta^{\smash{f}}_X|_A(\mu_1,\mu_2) &= \sup \left\{  \sum_{i \in I} F_f\ {\mu_1(A_i)}\ {\mu_2(A_i)}     \middle| 
\begin{array}{l@{}}
\{A_i\}_{i\in I} : \\
\text{fin. m'ble part. of }A
\end{array}
\right\}
\end{align*}
We write $\overline{\Delta}^{\smash{f}}_X|_A(\mu_1,\mu_2)$ for the right-hand side, and similary we can define $\overline{\Delta}^{\smash{f}}_X(\mu_1,\mu_2)$ . 
We formalize the above equality.
\begin{isabelle}
\isakeywordONE{definition}\isamarkupfalse%
\ f{\isacharunderscore}{\kern0pt}div{\isacharunderscore}{\kern0pt}pos{\isacharunderscore}{\kern0pt}sub{\isacharunderscore}{\kern0pt}partition\ \isakeywordTWO{where}\isanewline
\ \ {\isachardoublequoteopen}f{\isacharunderscore}{\kern0pt}div{\isacharunderscore}{\kern0pt}pos{\isacharunderscore}{\kern0pt}sub{\isacharunderscore}{\kern0pt}partition\ f\ L\ M\ N\ A\isanewline
\ {\isacharequal}\ Sup\ {\isacharbraceleft}{\kern0pt}{\isacharparenleft}{\kern0pt}{\isasymSum}\ C\ {\isasymin}\ S{\isachardot}{\kern0pt}\ e{\isadigit{2}}ennreal\ {\isacharparenleft}{\kern0pt}{\isacharparenleft}{\kern0pt}extend{\isacharunderscore}{\kern0pt}frac{\isacharunderscore}{\kern0pt}F\ f{\isacharparenright}{\kern0pt}\ {\isacharparenleft}{\kern0pt}measure\ M\ C{\isacharparenright}{\kern0pt}\ {\isacharparenleft}{\kern0pt}measure\ N\ C{\isacharparenright}{\kern0pt}{\isacharparenright}{\kern0pt}{\isacharparenright}{\kern0pt}\ {\isacharbar}{\kern0pt}\ S{\isachardot}{\kern0pt}\ S\ {\isasymin}\ measurable{\isacharunderscore}{\kern0pt}partition\ L\ A\ {\isacharbraceright}{\kern0pt}{\isachardoublequoteclose}
\end{isabelle}
\begin{isabelle}
\isakeywordONE{lemma}\isamarkupfalse%
\ {\isacharparenleft}{\kern0pt}\isakeywordTWO{in}\ densities{\isacharunderscore}{\kern0pt}two{\isacharunderscore}{\kern0pt}measures{\isacharunderscore}{\kern0pt}finite{\isacharparenright}{\kern0pt}\isanewline
f{\isacharunderscore}{\kern0pt}div{\isacharunderscore}{\kern0pt}pos{\isacharunderscore}{\kern0pt}sub{\isacharunderscore}{\kern0pt}eq{\isacharunderscore}{\kern0pt}f{\isacharunderscore}{\kern0pt}div{\isacharunderscore}{\kern0pt}pos{\isacharunderscore}{\kern0pt}sub{\isacharunderscore}{\kern0pt}partition{\isacharcolon}{\kern0pt}\isanewline
\ \ \isakeywordTWO{fixes}\ f{\isacharcolon}{\kern0pt}{\isacharcolon}{\kern0pt}{\isachardoublequoteopen}real\ {\isasymRightarrow}\ real{\isachardoublequoteclose}\ \isakeywordTWO{assumes}\ A{\isacharbrackleft}{\kern0pt}measurable{\isacharbrackright}{\kern0pt}{\isacharcolon}{\kern0pt}\ {\isachardoublequoteopen}A\ {\isasymin}\ sets\ L{\isachardoublequoteclose}\isanewline
\ \ \ \ \isakeywordTWO{and}\ gf{\isacharcolon}{\kern0pt}\ {\isachardoublequoteopen}generator{\isacharunderscore}{\kern0pt}F\ f{\isachardoublequoteclose}\ \isakeywordTWO{and}\ posf{\isacharcolon}{\kern0pt}\ {\isachardoublequoteopen}{\isasymforall}x\ {\isachargreater}{\kern0pt}\ {\isadigit{0}}{\isachardot}{\kern0pt}\ f\ x\ {\isasymge}\ {\isadigit{0}}{\isachardoublequoteclose}\isanewline
\ \ \isakeywordTWO{shows}\ {\isachardoublequoteopen}f{\isacharunderscore}{\kern0pt}div{\isacharunderscore}{\kern0pt}pos{\isacharunderscore}{\kern0pt}sub{\isacharunderscore}{\kern0pt}partition\ f\ L\ M\ N\ A\ {\isacharequal}{\kern0pt}\ f{\isacharunderscore}{\kern0pt}div{\isacharunderscore}{\kern0pt}pos{\isacharunderscore}{\kern0pt}sub\ f\ L\ M\ N\ A{\isachardoublequoteclose}
\end{isabelle}
\paragraph{A sketch of formal proof}
The formal proof is almost an interpretation of the proof of \cite[Thm. 4]{DBLP:conf/lics/SatoBGHK19}.
It is combining the piecewise monotonicity of $f$ and the following facts.
\begin{lemma}\label{lem:continuity:contableadd:bar}
 $\sum_{i \in I} \overline{\Delta}^{\smash{f}}_X|_{A_i}(\mu_1,\mu_2) = \overline{\Delta}^{\smash{f}}_X|_{(\bigcup_{i \in I} A_i)}(\mu_1,\mu_2)$ holds for countable disjoint family of measurable sets $\{A_i\}_{i \in I}$.
\end{lemma}
\begin{lemma}
$\overline{\Delta}^{\smash{f}}_X|_A= \Delta^{\smash{f}}_X|_A$ if
$A \subseteq p_1\inverse (0)$ or $A \subseteq p_2\inverse (0)$.
\end{lemma}
\begin{lemma}
If $a,b,c,d \in\Real$ and $C \in \Sigma_X$ satisfies 
$0 < a$,
$0 \leq c$,
$\forall x \in C.  p_1(x)/p_2(x) \in [a,b] \land f(p_1(x)/p_2(x)) \in [c,d]$ and
$f$ is either monotone increasing or decreasing on $[a,b]$ then
$
\Delta^{\smash{f}}_X|_C(\mu_1,\mu_2)
\leq \overline{\Delta}^{\smash{f}}_X|_C(\mu_1,\mu_2)  \leq \Delta^{\smash{f}}_X|_C(\mu_1,\mu_2) + (d-c)\mu_2(C)
$.
\end{lemma}
The overall formal proof spans about 3,300 lines,
including formalization of measurable partitions, $\overline{\Delta}^{\smash{f}}_X|_A$ and exchange
between integral and countable sums written in \isa{infsum}.

We also formalize the following trivial equality
\[
\overline{\Delta}^{\smash{f}}_X(\mu_1,\mu_2) = \sup_{\substack {I:\text{finite}, h \colon X \to I}} \Delta^{\smash{f}}_I(\prob(h)  (\mu_1),\prob(h) (\mu_2))
\]
where $\prob(h) (\mu_k) $ is $\mu_k \bind (\lambda x. \return h (x))$ ($k = 1,2$).
Due to the type system of Isabelle/HOL, we restrict $I$ to subset of $\Nat$.

\subsection{Approximation on two probabity kernels}

To prove the composition theorem of RDP, 
density functions of $c_1(x)$ and $c_2 (x)$ for $h, k \colon X \to \prob (Y)$ and $x \in X$ might be problematic.
For example if $X = Y = \Real$ and $h = \return$, there is no $\sigma$-finite measue dominating all $h(x)$ ($x \in \Real$).

To avoid this problem, we approximate $h, k \colon X \to \prob (Y)$. 
From $\Delta^{\smash{f}}_Y(\nu_1,\nu_2) = \sup_{I:\text{finite}, h \colon Y \to I} \Delta^{\smash{f}}_I(h (\nu_1),h(\nu_2))$,
$Y$ can be reduced to finite setting.
To reduce the domain $X$, the study \cite{DBLP:conf/lics/SatoBGHK19} shows the following property on $f$-divergence, called \emph{
approximability}.
\begin{lemma}
Let $h,k \colon X \to \prob (I)$ where $I$ is finite.
There is a family of finite sets $J_i$ and measurable functions $n_i \colon X \to J_i$ and $m_i \colon J_i \to X$ ($i \in \Nat$) such that
\begin{align*}
&\Delta^{\smash{f}}_I(\mu_1\bind (h \circ m_i \circ n_i),\mu_2 \bind (k \circ m_i \circ n_i))\\
&\quad \longrightarrow \Delta^{\smash{f}}_I (\mu_1\bind h,\mu_2 \bind k)\quad \text{as}\quad  i  \longrightarrow \infty
\end{align*}
\end{lemma}
Again, due to the type system of Isabelle/HOL, we adjust each $J_i$ to subset of $\Nat$ when we formalize this lemma.
\begin{isabelle}
\isakeywordONE{lemma}\isamarkupfalse%
\ f{\isacharunderscore}{\kern0pt}div{\isacharunderscore}{\kern0pt}pos{\isacharunderscore}{\kern0pt}sub{\isacharunderscore}{\kern0pt}approximable{\isacharcolon}{\kern0pt}\isanewline
\ \ \isakeywordTWO{fixes}\ h\ k{\isacharcolon}{\kern0pt}{\isacharcolon}{\kern0pt}{\isachardoublequoteopen}{\isacharprime}{\kern0pt}a\ {\isasymRightarrow}\ nat\ measure{\isachardoublequoteclose}\ \isakeywordTWO{and}\ X\ M\ N{\isacharcolon}{\kern0pt}{\isacharcolon}{\kern0pt}{\isachardoublequoteopen}{\isacharprime}{\kern0pt}a\ measure{\isachardoublequoteclose}\ \isakeywordTWO{and}\ I{\isacharcolon}{\kern0pt}{\isacharcolon}{\kern0pt}{\isachardoublequoteopen}nat\ set{\isachardoublequoteclose}\isanewline
\ \ \isakeywordTWO{assumes}\ finI{\isacharcolon}{\kern0pt}\ {\isachardoublequoteopen}finite\ I{\isachardoublequoteclose}\ \isakeywordTWO{and}\ A{\isacharcolon}{\kern0pt}\ {\isachardoublequoteopen}A\ {\isasymsubseteq}\ I{\isachardoublequoteclose}\isanewline
\ \ \ \ \isakeywordTWO{and}\ gf{\isacharcolon}{\kern0pt}\ {\isachardoublequoteopen}generator{\isacharunderscore}{\kern0pt}F\ f{\isachardoublequoteclose}\ \isakeywordTWO{and}\ posf{\isacharcolon}{\kern0pt}\ {\isachardoublequoteopen}{\isasymforall}x\ {\isachargreater}{\kern0pt}\ {\isadigit{0}}{\isachardot}{\kern0pt}\ {\isadigit{0}}\ {\isasymle}\ f\ x{\isachardoublequoteclose}\isanewline
\ \ \ \ \isakeywordTWO{and}\ meas{\isacharbrackleft}{\kern0pt}measurable{\isacharbrackright}{\kern0pt}{\isacharcolon}{\kern0pt}\ {\isachardoublequoteopen}h\ {\isasymin}\ X\ {\isasymrightarrow}\isactrlsub M\ \ subprob{\isacharunderscore}{\kern0pt}algebra\ {\isacharparenleft}{\kern0pt}count{\isacharunderscore}{\kern0pt}space\ I{\isacharparenright}{\kern0pt}{\isachardoublequoteclose}\ {\isachardoublequoteopen}k\ {\isasymin}\ X\ {\isasymrightarrow}\isactrlsub M\ \ subprob{\isacharunderscore}{\kern0pt}algebra\ {\isacharparenleft}{\kern0pt}count{\isacharunderscore}{\kern0pt}space\ I{\isacharparenright}{\kern0pt}{\isachardoublequoteclose}\ \isakeywordTWO{and}\ subprob{\isacharcolon}{\kern0pt}\ {\isachardoublequoteopen}M\ {\isasymin}\ space\ {\isacharparenleft}{\kern0pt}subprob{\isacharunderscore}{\kern0pt}algebra\ X{\isacharparenright}{\kern0pt}{\isachardoublequoteclose}\ {\isachardoublequoteopen}N\ {\isasymin}\ space\ {\isacharparenleft}{\kern0pt}subprob{\isacharunderscore}{\kern0pt}algebra\ X{\isacharparenright}{\kern0pt}{\isachardoublequoteclose}\isanewline
\ \ \isakeywordTWO{shows}\ {\isachardoublequoteopen}{\isasymexists}J{\isacharcolon}{\kern0pt}{\isacharcolon}{\kern0pt}nat\ {\isasymRightarrow}\ nat\ set{\isachardot}{\kern0pt}\isanewline
\ \ \ \ \ \ {\isasymexists}n{\isacharcolon}{\kern0pt}{\isacharcolon}{\kern0pt}nat\ {\isasymRightarrow}\ {\isacharprime}{\kern0pt}a\ {\isasymRightarrow}\ nat{\isachardot}{\kern0pt}\ {\isasymexists}m{\isacharcolon}{\kern0pt}{\isacharcolon}{\kern0pt}nat\ {\isasymRightarrow}\ {\isacharparenleft}{\kern0pt}nat\ {\isasymRightarrow}\ {\isacharprime}{\kern0pt}a{\isacharparenright}{\kern0pt}{\isachardot}{\kern0pt}\isanewline
\ \ \ \ \ \ {\isacharparenleft}{\kern0pt}{\isasymforall}i{\isachardot}{\kern0pt}\ finite\ {\isacharparenleft}{\kern0pt}J\ i{\isacharparenright}{\kern0pt}{\isacharparenright}{\kern0pt}\ {\isasymand}\ {\isacharparenleft}{\kern0pt}{\isasymforall}i{\isachardot}{\kern0pt}\ {\isacharparenleft}{\kern0pt}n\ i{\isacharparenright}{\kern0pt}\ {\isasymin}\ \ X\ {\isasymrightarrow}\isactrlsub M\ {\isacharparenleft}{\kern0pt}count{\isacharunderscore}{\kern0pt}space\ {\isacharparenleft}{\kern0pt}J\ i{\isacharparenright}{\kern0pt}{\isacharparenright}{\kern0pt}{\isacharparenright}{\kern0pt}\isanewline
\ \ \ \ \ \ {\isasymand}\ {\isacharparenleft}{\kern0pt}{\isasymforall}i{\isachardot}{\kern0pt}\ {\isacharparenleft}{\kern0pt}m\ i{\isacharparenright}{\kern0pt}\ {\isasymin}\ \ {\isacharparenleft}{\kern0pt}count{\isacharunderscore}{\kern0pt}space\ {\isacharparenleft}{\kern0pt}J\ i{\isacharparenright}{\kern0pt}{\isacharparenright}{\kern0pt}\ {\isasymrightarrow}\isactrlsub M\ X{\isacharparenright}{\kern0pt}\isanewline
\ \ \ \ \ \ {\isasymand}\ {\isacharparenleft}{\kern0pt}{\isacharparenleft}{\kern0pt}{\isasymlambda}i{\isacharcolon}{\kern0pt}{\isacharcolon}{\kern0pt}nat{\isachardot}{\kern0pt}\ \ {\isacharparenleft}{\kern0pt}f{\isacharunderscore}{\kern0pt}div{\isacharunderscore}{\kern0pt}pos{\isacharunderscore}{\kern0pt}sub\ f\ {\isacharparenleft}{\kern0pt}count{\isacharunderscore}{\kern0pt}space\ I{\isacharparenright}{\kern0pt}\ {\isacharparenleft}{\kern0pt}M{\isasymbind}{\isacharparenleft}{\kern0pt}{\isasymlambda}x{\isachardot}{\kern0pt}\ h{\isacharparenleft}{\kern0pt}{\isacharparenleft}{\kern0pt}m\ i{\isacharparenright}{\kern0pt}{\isacharparenleft}{\kern0pt}n\ i\ x{\isacharparenright}{\kern0pt}{\isacharparenright}{\kern0pt}{\isacharparenright}{\kern0pt}{\isacharparenright}{\kern0pt}\ {\isacharparenleft}{\kern0pt}N{\isasymbind}{\isacharparenleft}{\kern0pt}{\isasymlambda}x{\isachardot}{\kern0pt}\ k{\isacharparenleft}{\kern0pt}{\isacharparenleft}{\kern0pt}m\ i{\isacharparenright}{\kern0pt}{\isacharparenleft}{\kern0pt}n\ i\ x{\isacharparenright}{\kern0pt}{\isacharparenright}{\kern0pt}{\isacharparenright}{\kern0pt}{\isacharparenright}{\kern0pt}{\isacharparenright}{\kern0pt}\ A{\isacharparenright}{\kern0pt}\ {\isasymlonglonglongrightarrow}\ {\isacharparenleft}{\kern0pt}f{\isacharunderscore}{\kern0pt}div{\isacharunderscore}{\kern0pt}pos{\isacharunderscore}{\kern0pt}sub\ f\ {\isacharparenleft}{\kern0pt}count{\isacharunderscore}{\kern0pt}space\ I{\isacharparenright}{\kern0pt}\ {\isacharparenleft}{\kern0pt}M{\isasymbind}h{\isacharparenright}{\kern0pt}\ {\isacharparenleft}{\kern0pt}N{\isasymbind}k{\isacharparenright}{\kern0pt}{\isacharparenright}{\kern0pt}\ A{\isacharparenright}{\kern0pt}{\isachardoublequoteclose}
\end{isabelle}
\paragraph{A sketch of formal proof}
We adjust the proof in \cite{DBLP:conf/lics/SatoBGHK19} to the type system of Isabelle/HOL.
First, for measurable functions $h,k \colon X \to [0,R]$ with $R \in (0,\infty)$, and given $0 < \varepsilon$,
we define $J \subseteq \Nat \times \Nat$ and measurable $n \colon X \to J$ and $m \colon J \to X$ by
$n (x) = (\lceil h(x)/\varepsilon \rceil,\lceil k(x)/\varepsilon \rceil)$, $J = n(X)$ and $m (t_1,t_2) \in n\inverse (t_1,t_2)$.
By the definition of $n$,
$h(m(n(x))) = 0 \Leftrightarrow h(x) = 0$ and $k(m (n(x))) = 0  \Leftrightarrow k(x) = 0$ hold.

Second, we extend this result to a family $h_i,k_i \colon X \to [0,R]$ ($i \in I$) for finite $I \subseteq \Nat$.
Intuitively we take a product $J \simeq \prod_{i \in I} J_i$ where each $J_i$ is given by the first step for $h_i, k_i$.

Third, measurable functions $h,k \colon X \to \prob (I)$ can be converted to a family 
$(\lambda x. h(x)(j)),(\lambda x. k(x)(j)) \colon X \to [0,1]$  ($j \in I$).
By applying the second construction with $\varepsilon = 1/(i+1)$ for each $i \in \Nat$, we obtain $(J_i,n_i,m_i)$ such that 
$(f \circ m_i \circ n_i)$ and $(g \circ m_i \circ n_i)$ 
converge uniformly to $f$ and $g$ respectively. 
We then convert each $J_i \subseteq (I \to \Nat \times \Nat) $  to $J'_i \subseteq \Nat$ and modify $m_i$ and $n_i$ as expected.

Fourth, we show that for each $A \subseteq I$,
$(\mu_1\bind(f \circ m_i \circ n_i))(A)$ and $(\mu_2\bind(g \circ m_i \circ n_i))(A)$ converge to $(\mu_1\bind f)(A)$ and $(\mu_2\bind g)(A)$ respectively.
We then conclude the main statement easily, because if these limits are $0$ then the limiting sequences is $0,0,\ldots$ by construction of $n_i$. 

The overall formal proof spans about 1,500 lines.
\paragraph{Postprocessing inequality}
Postprocessing inequality is formalized from the ones of approximabiliy, continuity, and (finite) Jensen inequality for extension $F_f$.
\begin{isabelle}
\isakeywordONE{lemma}\isamarkupfalse%
\ f{\isacharunderscore}{\kern0pt}div{\isacharunderscore}{\kern0pt}pos{\isacharunderscore}{\kern0pt}postprocessing{\isacharcolon}{\kern0pt}\isanewline
\ \ \isakeywordTWO{assumes}\ gf{\isacharcolon}{\kern0pt}\ {\isachardoublequoteopen}generator{\isacharunderscore}{\kern0pt}F\ f{\isachardoublequoteclose}\isanewline
\ \ \ \ \isakeywordTWO{and}\ posf{\isacharcolon}{\kern0pt}\ {\isachardoublequoteopen}{\isasymforall}x\ {\isachargreater}{\kern0pt}\ {\isadigit{0}}{\isachardot}{\kern0pt}\ {\isadigit{0}}\ {\isasymle}\ f\ x{\isachardoublequoteclose}\isanewline
\ \ \ \ \isakeywordTWO{and}\ h{\isacharbrackleft}{\kern0pt}measurable{\isacharbrackright}{\kern0pt}{\isacharcolon}{\kern0pt}\ {\isachardoublequoteopen}h\ {\isasymin}\ L\ {\isasymrightarrow}\isactrlsub M\ \ subprob{\isacharunderscore}{\kern0pt}algebra\ K{\isachardoublequoteclose}\isanewline
\ \ \ \ \isakeywordTWO{and}\ cont{\isadigit{1}}{\isacharcolon}{\kern0pt}\ {\isachardoublequoteopen}densities{\isacharunderscore}{\kern0pt}two{\isacharunderscore}{\kern0pt}measures{\isacharunderscore}{\kern0pt}subprobabilistic\ L\ M\ N{\isachardoublequoteclose}\isanewline
\ \ \ \ \isakeywordTWO{and}\ cont{\isadigit{2}}{\isacharcolon}{\kern0pt}\ {\isachardoublequoteopen}densities{\isacharunderscore}{\kern0pt}two{\isacharunderscore}{\kern0pt}measures\ K\ {\isacharparenleft}{\kern0pt}M\ {\isasymbind}\ h{\isacharparenright}{\kern0pt}\ {\isacharparenleft}{\kern0pt}N\ {\isasymbind}\ h{\isacharparenright}{\kern0pt}{\isachardoublequoteclose}\isanewline
\ \ \isakeywordTWO{shows}\ {\isachardoublequoteopen}f{\isacharunderscore}{\kern0pt}div{\isacharunderscore}{\kern0pt}pos\ f\ K\ {\isacharparenleft}{\kern0pt}M{\isasymbind}h{\isacharparenright}{\kern0pt}{\isacharparenleft}{\kern0pt}N{\isasymbind}h{\isacharparenright}{\kern0pt}\ {\isasymle}\ f{\isacharunderscore}{\kern0pt}div{\isacharunderscore}{\kern0pt}pos\ f\ L\ M\ N{\isachardoublequoteclose}
\end{isabelle}
We remark here that \isa{K} does not need to dominate each (\isa{h x}).
It suffices to dominate only two measures \isa{M\ {\isasymbind}\ h} and \isa{N\ {\isasymbind}\ h}.
This implies that the formalization supports general settings: we may replace base measures by sums of appropriate measures.

\section{RDP in Isabelle/HOL}
\subsection{R\'{e}nyi divergence}

We formalize R\'{e}nyi divergence, based on the reformulation 
\[
D^\alpha_X(\mu_1,\mu_2) = \frac{1}{\alpha - 1}\log \Delta^{f_\alpha}_X(\mu_1,\mu_2) \qquad (1<\alpha).
\] 
We recall that the generator function is $f_\alpha(t) = t^\alpha$.
\begin{isabelle}
\isakeywordONE{definition}\isamarkupfalse%
\ Renyi{\isacharunderscore}{\kern0pt}divergence\ \isakeywordTWO{where}\isanewline
\ \ {\isachardoublequoteopen}Renyi{\isacharunderscore}{\kern0pt}divergence\ {\isasymalpha}\ L\ M\ N\isanewline
\ \ {\isacharequal}{\kern0pt}\ Renyi{\isacharunderscore}{\kern0pt}log{\isacharunderscore}{\kern0pt}part\ {\isasymalpha}\ {\isacharparenleft}{\kern0pt}f{\isacharunderscore}{\kern0pt}div{\isacharunderscore}{\kern0pt}pos\ {\isacharparenleft}{\kern0pt}{\isasymlambda}t{\isachardot}{\kern0pt}\ \ t\ powr\ {\isasymalpha}{\isacharparenright}{\kern0pt}\ L\ M\ N{\isacharparenright}{\kern0pt}{\isachardoublequoteclose}
\end{isabelle}
\begin{isabelle}
\isakeywordONE{abbreviation}\isamarkupfalse%
\ {\isachardoublequoteopen}Renyi{\isacharunderscore}{\kern0pt}log{\isacharunderscore}{\kern0pt}part\ {\isasymalpha}\ {\isasymequiv}\ \ {\isacharparenleft}{\kern0pt}{\isasymlambda}z{\isacharcolon}{\kern0pt}{\isacharcolon}{\kern0pt}ennreal{\isachardot}{\kern0pt}\ if\ {\isacharparenleft}{\kern0pt}z\ {\isacharequal}{\kern0pt}\ {\isadigit{0}}{\isacharparenright}{\kern0pt}\ then\ {\isacharminus}{\kern0pt}{\isasyminfinity}\ else\ if\ {\isacharparenleft}{\kern0pt}z\ {\isacharless}{\kern0pt}\ {\isasyminfinity}{\isacharparenright}{\kern0pt}\ then\ ereal{\isacharparenleft}{\kern0pt}ln\ {\isacharparenleft}{\kern0pt}enn{\isadigit{2}}real\ z{\isacharparenright}{\kern0pt}\ {\isacharslash}{\kern0pt}\ {\isacharparenleft}{\kern0pt}{\isasymalpha}{\isacharminus}{\kern0pt}{\isadigit{1}}{\isacharparenright}{\kern0pt}{\isacharparenright}{\kern0pt}\ else\ {\isasyminfinity}{\isacharparenright}{\kern0pt}{\isachardoublequoteclose}
\end{isabelle}
Here, \isa{Renyi{\isacharunderscore}{\kern0pt}log{\isacharunderscore}{\kern0pt}part\ {\isasymalpha}} is the mapping 
$t \mapsto \frac{1}{\alpha - 1}\log(t)$.
In Isabelle's standard formalization, $\log$ is defined only for real numbers and set by $\log(0) = 0$ at $0$.
To formalize R\'{e}nyi divergence, we need to introduce the extension $\frac{1}{\alpha - 1}\log(\infty) = \infty$.
We also introduce $\frac{1}{\alpha - 1}\log(0) = -\infty$ to capture R\'{e}nyi divergence between subprobability measures, and it is convenient that the mapping $t \mapsto \frac{1}{\alpha - 1}\log(t)$ is then monotone and bijective. The inverse is,
\begin{isabelle}
\isakeywordONE{abbreviation}\isamarkupfalse%
\ {\isachardoublequoteopen}inv{\isacharunderscore}{\kern0pt}Renyi{\isacharunderscore}{\kern0pt}log{\isacharunderscore}{\kern0pt}part\ {\isasymalpha}\ {\isasymequiv}\ {\isacharparenleft}{\kern0pt}{\isasymlambda}z{\isacharcolon}{\kern0pt}{\isacharcolon}{\kern0pt}ereal{\isachardot}{\kern0pt}\ if\ {\isacharparenleft}{\kern0pt}z\ {\isacharequal}{\kern0pt}\ {\isacharminus}{\kern0pt}{\isasyminfinity}{\isacharparenright}{\kern0pt}\ then\ {\isadigit{0}}\ else\ if\ {\isacharparenleft}{\kern0pt}z\ {\isacharless}{\kern0pt}\ {\isasyminfinity}{\isacharparenright}{\kern0pt}\ then\ ennreal{\isacharparenleft}{\kern0pt}exp{\isacharparenleft}{\kern0pt}{\isacharparenleft}{\kern0pt}real{\isacharunderscore}{\kern0pt}of{\isacharunderscore}{\kern0pt}ereal\ z{\isacharparenright}{\kern0pt}\ {\isacharasterisk}{\kern0pt}\ {\isacharparenleft}{\kern0pt}{\isasymalpha}{\isacharminus}{\kern0pt}{\isadigit{1}}{\isacharparenright}{\kern0pt}{\isacharparenright}{\kern0pt}{\isacharparenright}{\kern0pt}\ else\ {\isasyminfinity}{\isacharparenright}{\kern0pt}{\isachardoublequoteclose}
\end{isabelle}
From the well-definedness of $\Delta^{\smash{f^\alpha}}_X$, the base measure is dropped from the formalization:
\begin{isabelle}
\isakeywordONE{abbreviation}\isamarkupfalse%
\ {\isachardoublequoteopen}Renyi{\isacharunderscore}{\kern0pt}div\ {\isasymequiv}\ {\isacharparenleft}{\kern0pt}{\isasymlambda}{\isasymalpha}\ M\ N{\isachardot}{\kern0pt}\ Renyi{\isacharunderscore}{\kern0pt}divergence\ {\isasymalpha}\ {\isacharparenleft}{\kern0pt}sum{\isacharunderscore}{\kern0pt}measure\ M\ N{\isacharparenright}{\kern0pt}\ M\ N\ {\isacharparenright}{\kern0pt}{\isachardoublequoteclose}
\end{isabelle}

\subsection{RDP in Isabelle/HOL}
We now fomalize the predicates for RDP in the same way as 
the formalization of DP~\cite{10.1145/3703595.3705875,Differential_Privacy-AFP}.
\begin{isabelle}
\isakeywordONE{definition}\isamarkupfalse%
\ Renyi{\isacharunderscore}{\kern0pt}differential{\isacharunderscore}{\kern0pt}privacy\ {\isacharcolon}{\kern0pt}{\isacharcolon}{\kern0pt}\ {\isachardoublequoteopen}{\isacharparenleft}{\kern0pt}{\isacharprime}{\kern0pt}a\ {\isasymRightarrow}\ {\isacharprime}{\kern0pt}b\ measure{\isacharparenright}{\kern0pt}\ {\isasymRightarrow}\ {\isacharparenleft}{\kern0pt}{\isacharprime}{\kern0pt}a\ rel{\isacharparenright}{\kern0pt}\ {\isasymRightarrow}\ real\ {\isasymRightarrow}\ real\ {\isasymRightarrow}\ bool\ {\isachardoublequoteclose}\isakeywordTWO{where}\isanewline
\ \ {\isachardoublequoteopen}Renyi{\isacharunderscore}{\kern0pt}differential{\isacharunderscore}{\kern0pt}privacy\ M\ adj\ {\isasymalpha}\ r\ {\isasymequiv}\ {\isasymforall}{\isacharparenleft}{\kern0pt}d{\isadigit{1}}{\isacharcomma}{\kern0pt}d{\isadigit{2}}{\isacharparenright}{\kern0pt}{\isasymin}adj{\isachardot}{\kern0pt}\ Renyi{\isacharunderscore}{\kern0pt}div\ {\isasymalpha}\ {\isacharparenleft}{\kern0pt}M\ d{\isadigit{1}}{\isacharparenright}{\kern0pt}\ {\isacharparenleft}{\kern0pt}M\ d{\isadigit{2}}{\isacharparenright}{\kern0pt}\ {\isasymle}\ r\ {\isasymand}\ Renyi{\isacharunderscore}{\kern0pt}div\ {\isasymalpha}\ {\isacharparenleft}{\kern0pt}M\ d{\isadigit{2}}{\isacharparenright}{\kern0pt}\ {\isacharparenleft}{\kern0pt}M\ d{\isadigit{1}}{\isacharparenright}{\kern0pt}\ {\isasymle}\ r{\isachardoublequoteclose}
\end{isabelle}

\paragraph{Basic properties}
We formalize basic properties: zero property, postprocessing inequality (Lemma \ref{RDP:postprocess}), and composition theorem (Lemma \ref{RDP:compose:adaptive}).
For example we formalize the composition theorem as follows:
\begin{isabelle}
\isakeywordONE{proposition}\isamarkupfalse%
\ Renyi{\isacharunderscore}{\kern0pt}differential{\isacharunderscore}{\kern0pt}privacy{\isacharunderscore}{\kern0pt}composition{\isacharunderscore}{\kern0pt}adaptive{\isacharcolon}{\kern0pt}\isanewline
\ \ \isakeywordTWO{assumes}\ {\isachardoublequoteopen}{\isadigit{1}}\ {\isacharless}{\kern0pt}\ {\isasymalpha}{\isachardoublequoteclose}\ \isakeywordTWO{and}\ measM{\isacharcolon}{\kern0pt}\ {\isachardoublequoteopen}M\ {\isasymin}\ X\ {\isasymrightarrow}\isactrlsub M\ {\isacharparenleft}{\kern0pt}subprob{\isacharunderscore}{\kern0pt}algebra\ Y{\isacharparenright}{\kern0pt}{\isachardoublequoteclose}\isanewline
\ \ \ \ \isakeywordTWO{and}\ RDPM{\isacharcolon}{\kern0pt}\ {\isachardoublequoteopen}Renyi{\isacharunderscore}{\kern0pt}differential{\isacharunderscore}{\kern0pt}privacy\ M\ adj\ {\isasymalpha}\ r{\isachardoublequoteclose}\isanewline
\ \ \ \ \isakeywordTWO{and}\ measN{\isacharcolon}{\kern0pt}\ {\isachardoublequoteopen}N\ {\isasymin}\ {\isacharparenleft}{\kern0pt}X\ {\isasymOtimes}\isactrlsub M\ Y{\isacharparenright}{\kern0pt}\ {\isasymrightarrow}\isactrlsub M\ {\isacharparenleft}{\kern0pt}subprob{\isacharunderscore}{\kern0pt}algebra\ Z{\isacharparenright}{\kern0pt}{\isachardoublequoteclose}\isanewline
\ \ \ \ \isakeywordTWO{and}\ RDPN{\isacharcolon}{\kern0pt}\ {\isachardoublequoteopen}{\isasymforall}y\ {\isasymin}\ space\ Y{\isachardot}{\kern0pt}\ Renyi{\isacharunderscore}{\kern0pt}differential{\isacharunderscore}{\kern0pt}privacy\ {\isacharparenleft}{\kern0pt}{\isasymlambda}\ x{\isachardot}{\kern0pt}\ N\ {\isacharparenleft}{\kern0pt}x{\isacharcomma}{\kern0pt}y{\isacharparenright}{\kern0pt}{\isacharparenright}{\kern0pt}\ adj\ {\isasymalpha}\ r{\isacharprime}{\kern0pt}{\isachardoublequoteclose}\isanewline
\ \ \ \ \isakeywordTWO{and}\ {\isachardoublequoteopen}adj\ {\isasymsubseteq}\ {\isacharparenleft}{\kern0pt}space\ X{\isacharparenright}{\kern0pt}\ {\isasymtimes}\ {\isacharparenleft}{\kern0pt}space\ X{\isacharparenright}{\kern0pt}{\isachardoublequoteclose}\isanewline
\ \ \isakeywordTWO{shows}\ {\isachardoublequoteopen}Renyi{\isacharunderscore}{\kern0pt}differential{\isacharunderscore}{\kern0pt}privacy\ {\isacharparenleft}{\kern0pt}{\isasymlambda}x{\isachardot}{\kern0pt}\ do{\isacharbraceleft}{\kern0pt}y\ {\isasymleftarrow}\ M\ x{\isacharsemicolon}{\kern0pt}\ N\ {\isacharparenleft}{\kern0pt}x{\isacharcomma}{\kern0pt}\ y{\isacharparenright}{\kern0pt}{\isacharbraceright}{\kern0pt}{\isacharparenright}{\kern0pt}\ adj\ {\isasymalpha}\ {\isacharparenleft}{\kern0pt}r\ {\isacharplus}{\kern0pt}\ r{\isacharprime}{\kern0pt}{\isacharparenright}{\kern0pt}{\isachardoublequoteclose}
\end{isabelle}
They are proved formalzation of the postprocessing inequality and the composability (Lemma \ref{Renyi:compose:monadic}).
The formal proof of the composability is based on the proof given in \cite{DBLP:conf/lics/SatoBGHK19}.
From the continuity and appriximablity of $f$-divergences, it suffices to formalize the composability of $\Delta^{f^\alpha}$ in the finite discrete setting. 

\subsection{Gaussian mechanism}
We formalize 1-dimentional Gaussian mechanism.
\begin{isabelle}
\isakeywordONE{definition}\isamarkupfalse%
\ Gaussian{\isacharunderscore}{\kern0pt}dist\ \ {\isacharcolon}{\kern0pt}{\isacharcolon}{\kern0pt}\ {\isachardoublequoteopen}real\ {\isasymRightarrow}\ real\ {\isasymRightarrow}\ real\ measure{\isachardoublequoteclose}\ \isakeywordTWO{where}\isanewline
\ \ {\isachardoublequoteopen}Gaussian{\isacharunderscore}{\kern0pt}dist\ {\isasymsigma}\ x\ {\isacharequal}{\kern0pt}\ {\isacharparenleft}{\kern0pt}if\ {\isasymsigma}\ {\isasymle}\ {\isadigit{0}}\ then\ return\ borel\ x\ else\ density\ lborel\ {\isacharparenleft}{\kern0pt}normal{\isacharunderscore}{\kern0pt}density\ x\ {\isasymsigma}{\isacharparenright}{\kern0pt}{\isacharparenright}{\kern0pt}{\isachardoublequoteclose}
\end{isabelle}
Here, \isa{density\ lborel\ {\isacharparenleft}{\kern0pt}normal{\isacharunderscore}{\kern0pt}density\ x\ {\isasymsigma}{\isacharparenright}} is a formalization of $\mathcal{N}(x,\sigma^2)$. 
To show the measurability of it, we apply the continuity of cumulative distribution functions formalized in the AFP entry \isa{Error\_Function} \cite{Error_Function-AFP}.
Formalization of its RDP is rather straightforward.
\begin{isabelle}
\isakeywordONE{lemma}\isamarkupfalse%
\ Renyi{\isacharunderscore}{\kern0pt}divergence{\isacharunderscore}{\kern0pt}Gaussian{\isacharunderscore}{\kern0pt}dist{\isacharcolon}{\kern0pt}\isanewline
\ \ \isakeywordTWO{fixes}\ x\ y\ {\isasymsigma}\ {\isacharcolon}{\kern0pt}{\isacharcolon}{\kern0pt}real
\ \isakeywordTWO{assumes}\ {\isacharbrackleft}{\kern0pt}arith{\isacharbrackright}{\kern0pt}{\isacharcolon}{\kern0pt}\ {\isachardoublequoteopen}{\isadigit{0}}\ {\isacharless}{\kern0pt}\ {\isasymsigma}{\isachardoublequoteclose}
\ \isakeywordTWO{and}\ a{\isacharcolon}{\kern0pt}\ {\isachardoublequoteopen}{\isadigit{1}}\ {\isacharless}{\kern0pt}\ {\isacharparenleft}{\kern0pt}{\isasymalpha}{\isacharcolon}{\kern0pt}{\isacharcolon}{\kern0pt}real{\isacharparenright}{\kern0pt}{\isachardoublequoteclose}\ \isanewline
\ \ \isakeywordTWO{shows}\ {\isachardoublequoteopen}Renyi{\isacharunderscore}{\kern0pt}divergence\ {\isasymalpha}\ lborel\ {\isacharparenleft}{\kern0pt}Gaussian{\isacharunderscore}{\kern0pt}dist\ {\isasymsigma}\ x{\isacharparenright}{\kern0pt}\ {\isacharparenleft}{\kern0pt}Gaussian{\isacharunderscore}{\kern0pt}dist\ {\isasymsigma}\ y{\isacharparenright}{\kern0pt}\ {\isacharequal}{\kern0pt}\ {\isacharparenleft}{\kern0pt}{\isasymalpha}\ {\isacharasterisk}{\kern0pt}\ {\isacharparenleft}{\kern0pt}x\ {\isacharminus}{\kern0pt}\ y{\isacharparenright}{\kern0pt}\isactrlsup {\isadigit{2}}\ {\isacharslash}{\kern0pt}\ {\isacharparenleft}{\kern0pt}{\isadigit{2}}\ {\isacharasterisk}{\kern0pt}\ {\isasymsigma}\isactrlsup {\isadigit{2}}{\isacharparenright}{\kern0pt}{\isacharparenright}{\kern0pt}{\isachardoublequoteclose}
\end{isabelle}

\paragraph{Multi-dimensional mechanism}

We formalize the multi-dimensional Gaussian mechanism 
$(\lambda \vec{x}. \mathcal{N}(\vec{x},\sigma^2 I )) \colon \Real^n \to \prob (\Real^n)$ by the following induction. 
\begin{isabelle}
\isakeywordONE{primrec}\isamarkupfalse%
\ Gaussian{\isacharunderscore}{\kern0pt}vec\ {\isacharcolon}{\kern0pt}{\isacharcolon}{\kern0pt}\ {\isachardoublequoteopen}nat\ {\isasymRightarrow}\ {\isacharparenleft}{\kern0pt}nat\ {\isasymRightarrow}\ real{\isacharparenright}{\kern0pt}\ {\isasymRightarrow}\ {\isacharparenleft}{\kern0pt}nat\ {\isasymRightarrow}\ real{\isacharparenright}{\kern0pt}\ measure{\isachardoublequoteclose}\ \isakeywordTWO{where}\isanewline
\ \ {\isachardoublequoteopen}Gaussian{\isacharunderscore}{\kern0pt}vec\ {\isadigit{0}}\ f\ {\isacharequal}{\kern0pt}\ \ return\ {\isacharparenleft}{\kern0pt}count{\isacharunderscore}{\kern0pt}space\ {\isacharbraceleft}{\kern0pt}{\isasymlambda}{\isacharunderscore}{\kern0pt}{\isachardot}{\kern0pt}\ undefined{\isacharbraceright}{\kern0pt}{\isacharparenright}{\kern0pt}\ {\isacharparenleft}{\kern0pt}{\isasymlambda}{\isacharunderscore}{\kern0pt}{\isachardot}{\kern0pt}\ undefined{\isacharparenright}{\kern0pt}{\isachardoublequoteclose}\isanewline
{\isacharbar}{\kern0pt}\ {\isachardoublequoteopen}Gaussian{\isacharunderscore}{\kern0pt}vec\ {\isacharparenleft}{\kern0pt}Suc\ k{\isacharparenright}{\kern0pt}\ f\ {\isacharequal}{\kern0pt}\ do{\isacharbraceleft}{\kern0pt}\ g\ {\isasymleftarrow}\ Gaussian{\isacharunderscore}{\kern0pt}vec\ k\ f{\isacharsemicolon}{\kern0pt}\ v\ {\isasymleftarrow}\ Gaussian{\isacharunderscore}{\kern0pt}dist\ {\isasymsigma}\ {\isacharparenleft}{\kern0pt}f\ k{\isacharparenright}{\kern0pt}{\isacharsemicolon}{\kern0pt}\ return\ {\isacharparenleft}{\kern0pt}{\isasymPi}\isactrlsub M\ x{\isasymin}{\isacharbraceleft}{\kern0pt}{\isadigit{0}}{\isachardot}{\kern0pt}{\isachardot}{\kern0pt}k{\isacharbraceright}{\kern0pt}{\isachardot}{\kern0pt}\ borel{\isacharparenright}{\kern0pt}\ {\isacharparenleft}{\kern0pt}g{\isacharparenleft}{\kern0pt}k\ {\isacharcolon}{\kern0pt}{\isacharequal}{\kern0pt}\ v{\isacharparenright}{\kern0pt}{\isacharparenright}{\kern0pt}{\isacharbraceright}{\kern0pt}{\isachardoublequoteclose}
\end{isabelle}
Formalization of its RDP is just iterating ones of 1-dimensional Gaussian mechanism. 
We omit it in this paper.

\subsection{Conversion to DP}

We formalize the three conversion laws from RDP to DP, 
given by Mironov \cite{Mironov17}(Lemma \ref{convRDPtoDP:Mironov}), Balle et atl. \cite{DBLP:journals/corr/abs-1905-09982}(Lemma \ref{convRDPtoDP:Balle}) and Asoodeh et al. \cite{DBLP:journals/corr/abs-2008-06529}(Lemma \ref{convRDPtoDP:Asoodeh}).
\begin{isabelle}
\isakeywordONE{lemma}\isamarkupfalse%
\ Renyi{\isacharunderscore}{\kern0pt}differential{\isacharunderscore}{\kern0pt}privacy{\isacharunderscore}{\kern0pt}to{\isacharunderscore}{\kern0pt}DP{\isacharcolon}{\kern0pt}\isanewline
\ \ \isakeywordTWO{assumes}\ {\isachardoublequoteopen}{\isadigit{1}}\ {\isacharless}{\kern0pt}\ {\isasymalpha}{\isachardoublequoteclose}\ \isakeywordTWO{and}\ RDPM{\isacharcolon}{\kern0pt}\ {\isachardoublequoteopen}Renyi{\isacharunderscore}{\kern0pt}differential{\isacharunderscore}{\kern0pt}privacy\ M\ adj\ {\isasymalpha}\ r{\isachardoublequoteclose}\isanewline
\ \ \ \ \isakeywordTWO{and}\ measM{\isacharbrackleft}{\kern0pt}measurable{\isacharbrackright}{\kern0pt}{\isacharcolon}{\kern0pt}\ {\isachardoublequoteopen}M\ {\isasymin}\ X\ {\isasymrightarrow}\isactrlsub M\ {\isacharparenleft}{\kern0pt}prob{\isacharunderscore}{\kern0pt}algebra\ Y{\isacharparenright}{\kern0pt}{\isachardoublequoteclose}\isanewline
\ \ \ \ \isakeywordTWO{and}\ adj{\isacharcolon}{\kern0pt}\ {\isachardoublequoteopen}adj\ {\isasymsubseteq}\ {\isacharparenleft}{\kern0pt}space\ X{\isacharparenright}{\kern0pt}\ {\isasymtimes}\ {\isacharparenleft}{\kern0pt}space\ X{\isacharparenright}{\kern0pt}{\isachardoublequoteclose}\ \isakeywordTWO{and}\ {\isachardoublequoteopen}{\isadigit{0}}\ {\isacharless}{\kern0pt}\ {\isasymdelta}{\isachardoublequoteclose}\ {\isachardoublequoteopen}{\isasymdelta}\ {\isacharless}{\kern0pt}\ {\isadigit{1}}{\isachardoublequoteclose}\isanewline
\ \ \isakeywordTWO{shows}\ {\isachardoublequoteopen}differential{\isacharunderscore}{\kern0pt}privacy\ M\ adj\ {\isacharparenleft}{\kern0pt}if\ {\isadigit{1}}\ {\isasymle}\ {\isasymalpha}\ {\isacharasterisk}{\kern0pt}\ {\isasymdelta}\ then\ max\ {\isadigit{0}}\ {\isacharparenleft}{\kern0pt}r\ {\isacharplus}{\kern0pt}\ ln\ {\isacharparenleft}{\kern0pt}{\isadigit{1}}\ {\isacharminus}{\kern0pt}\ {\isasymdelta}{\isacharparenright}{\kern0pt}{\isacharparenright}{\kern0pt}\ else\ min\ {\isacharparenleft}{\kern0pt}max\ {\isadigit{0}}\ {\isacharparenleft}{\kern0pt}r\ {\isacharplus}{\kern0pt}\ ln\ {\isacharparenleft}{\kern0pt}{\isacharparenleft}{\kern0pt}{\isasymalpha}\ {\isacharminus}{\kern0pt}\ {\isadigit{1}}{\isacharparenright}{\kern0pt}\ {\isacharslash}{\kern0pt}\ {\isasymalpha}{\isacharparenright}{\kern0pt}\ {\isacharminus}{\kern0pt}\ {\isacharparenleft}{\kern0pt}ln\ {\isasymdelta}\ {\isacharplus}{\kern0pt}\ ln\ {\isasymalpha}{\isacharparenright}{\kern0pt}\ {\isacharslash}{\kern0pt}\ {\isacharparenleft}{\kern0pt}{\isasymalpha}\ {\isacharminus}{\kern0pt}\ {\isadigit{1}}{\isacharparenright}{\kern0pt}{\isacharparenright}{\kern0pt}{\isacharparenright}{\kern0pt}\ {\isacharparenleft}{\kern0pt}ln\ {\isacharparenleft}{\kern0pt}{\isacharparenleft}{\kern0pt}exp\ {\isacharparenleft}{\kern0pt}r\ {\isacharasterisk}{\kern0pt}\ {\isacharparenleft}{\kern0pt}{\isasymalpha}\ {\isacharminus}{\kern0pt}\ {\isadigit{1}}{\isacharparenright}{\kern0pt}{\isacharparenright}{\kern0pt}\ {\isacharminus}{\kern0pt}\ {\isadigit{1}}{\isacharparenright}{\kern0pt}\ {\isacharslash}{\kern0pt}\ {\isacharparenleft}{\kern0pt}{\isasymalpha}\ {\isacharasterisk}{\kern0pt}\ {\isasymdelta}{\isacharparenright}{\kern0pt}\ {\isacharplus}{\kern0pt}\ {\isadigit{1}}{\isacharparenright}{\kern0pt}\ {\isacharslash}{\kern0pt}\ {\isacharparenleft}{\kern0pt}{\isasymalpha}\ {\isacharminus}{\kern0pt}\ {\isadigit{1}}{\isacharparenright}{\kern0pt}{\isacharparenright}{\kern0pt}{\isacharparenright}{\kern0pt}\ {\isasymdelta}{\isachardoublequoteclose}\isanewline
\ \ \ \ \isakeywordTWO{and}\ {\isachardoublequoteopen}differential{\isacharunderscore}{\kern0pt}privacy\ M\ adj\ {\isacharparenleft}{\kern0pt}max\ {\isadigit{0}}\ {\isacharparenleft}{\kern0pt}r\ {\isacharplus}{\kern0pt}\ ln\ {\isacharparenleft}{\kern0pt}{\isacharparenleft}{\kern0pt}{\isasymalpha}\ {\isacharminus}{\kern0pt}\ {\isadigit{1}}{\isacharparenright}{\kern0pt}\ {\isacharslash}{\kern0pt}\ {\isasymalpha}{\isacharparenright}{\kern0pt}\ {\isacharminus}{\kern0pt}\ {\isacharparenleft}{\kern0pt}ln\ {\isasymdelta}\ {\isacharplus}{\kern0pt}\ ln\ {\isasymalpha}{\isacharparenright}{\kern0pt}\ {\isacharslash}{\kern0pt}\ {\isacharparenleft}{\kern0pt}{\isasymalpha}\ {\isacharminus}{\kern0pt}\ {\isadigit{1}}{\isacharparenright}{\kern0pt}{\isacharparenright}{\kern0pt}{\isacharparenright}{\kern0pt}\ {\isasymdelta}{\isachardoublequoteclose}\isanewline
\ \ \ \ \isakeywordTWO{and}\ {\isachardoublequoteopen}differential{\isacharunderscore}{\kern0pt}privacy\ M\ adj\ {\isacharparenleft}{\kern0pt}r\ {\isacharminus}{\kern0pt}\ ln\ {\isasymdelta}\ {\isacharslash}{\kern0pt}\ {\isacharparenleft}{\kern0pt}{\isasymalpha}\ {\isacharminus}{\kern0pt}\ {\isadigit{1}}{\isacharparenright}{\kern0pt}{\isacharparenright}{\kern0pt}\ {\isasymdelta}{\isachardoublequoteclose}
\end{isabelle}
First two conversions are formalized based on the proofs given in \cite{DBLP:journals/corr/abs-1905-09982,ALCKS2021}.
We first formalize the following privacy regions (cf. \cite{KairouzOV15}) for divergences $\DP \varepsilon {}$ and $D^\alpha$.
%
%
\begin{align*}
R(\varepsilon,\delta) &= \left\{ (x,y) \in [0,1]^2 ~\middle|~ 
\begin{aligned}
x &\leq \exp(\varepsilon)(1-y) + \delta,\\
y &\leq \exp(\varepsilon)(1-x) + \delta  
\end{aligned}
\right\},\\
R(\alpha,\rho) &= \left\{ (x,y) \in [0,1]^2 ~\middle|~  
\begin{aligned}
&x^\alpha (1-y)^{1-\alpha} +  (1-x)^\alpha y^{1-\alpha} \\
&\leq \exp((\alpha-1) \rho)
\end{aligned}
\right\}.
\end{align*}
Since $0^{1-\alpha} = 0$ in Isabelle/HOL's standard library (we must have $0^{1-\alpha} = \infty$), 
we need case analyses with $y = 0,1$ to formalize $R(\alpha,\rho)$ .
\begin{isabelle}
\isakeywordONE{definition}\isamarkupfalse%
\ Region{\isacharunderscore}{\kern0pt}Renyi\ {\isacharcolon}{\kern0pt}{\isacharcolon}{\kern0pt}\ {\isachardoublequoteopen}real\ {\isasymRightarrow}\ real\ {\isasymRightarrow}\ {\isacharparenleft}{\kern0pt}real\ {\isasymtimes}\ real{\isacharparenright}{\kern0pt}\ set{\isachardoublequoteclose}\ \isakeywordTWO{where}\isanewline
\ \ {\isachardoublequoteopen}Region{\isacharunderscore}{\kern0pt}Renyi\ {\isasymalpha}\ r\ {\isacharequal}{\kern0pt}\ {\isacharbraceleft}{\kern0pt}{\isacharparenleft}{\kern0pt}x{\isacharcomma}{\kern0pt}y{\isacharparenright}{\kern0pt}\ {\isacharbar}{\kern0pt}x\ y{\isacharcolon}{\kern0pt}{\isacharcolon}{\kern0pt}real{\isachardot}{\kern0pt}\ \ {\isasymnot}{\isacharparenleft}{\kern0pt}y\ {\isacharequal}{\kern0pt}\ {\isadigit{1}}\ {\isasymand}\ {\isadigit{0}}\ {\isacharless}{\kern0pt}\ x\ {\isasymor}\ y\ {\isacharequal}{\kern0pt}\ {\isadigit{0}}\ {\isasymand}\ x\ {\isacharless}{\kern0pt}\ {\isadigit{1}}{\isacharparenright}{\kern0pt}\ {\isasymand}\ {\isacharparenleft}{\kern0pt}x\ powr\ {\isasymalpha}\ {\isacharasterisk}{\kern0pt}\ {\isacharparenleft}{\kern0pt}{\isadigit{1}}\ {\isacharminus}{\kern0pt}\ y{\isacharparenright}{\kern0pt}\ powr\ {\isacharparenleft}{\kern0pt}{\isadigit{1}}\ {\isacharminus}{\kern0pt}\ {\isasymalpha}{\isacharparenright}{\kern0pt}\ {\isacharplus}{\kern0pt}\ {\isacharparenleft}{\kern0pt}{\isadigit{1}}\ {\isacharminus}{\kern0pt}\ x{\isacharparenright}{\kern0pt}\ powr\ {\isasymalpha}\ {\isacharasterisk}{\kern0pt}\ y\ powr\ {\isacharparenleft}{\kern0pt}{\isadigit{1}}\ {\isacharminus}{\kern0pt}\ {\isasymalpha}{\isacharparenright}{\kern0pt}{\isacharparenright}{\kern0pt}\ {\isasymle}\ \ exp\ {\isacharparenleft}{\kern0pt}r\ {\isacharasterisk}{\kern0pt}\ {\isacharparenleft}{\kern0pt}{\isasymalpha}\ {\isacharminus}{\kern0pt}\ {\isadigit{1}}{\isacharparenright}{\kern0pt}{\isacharparenright}{\kern0pt}{\isacharbraceright}{\kern0pt}\ {\isasyminter}\ {\isacharbraceleft}{\kern0pt}{\isadigit{0}}{\isachardot}{\kern0pt}{\isachardot}{\kern0pt}{\isadigit{1}}{\isacharbraceright}{\kern0pt}{\isasymtimes}{\isacharbraceleft}{\kern0pt}{\isadigit{0}}{\isachardot}{\kern0pt}{\isachardot}{\kern0pt}{\isadigit{1}}{\isacharbraceright}{\kern0pt}{\isachardoublequoteclose}%
\end{isabelle}
We have a core of conversions (\cite[Sec. 5.1]{DBLP:journals/corr/abs-1905-09982}):  
\begin{align*}
&R(\alpha,\rho) \subseteq R(\varepsilon,\delta) \\
&\implies \forall X :\text{measurable space}, \forall \mu_1, \mu_2 \in \prob (X)\\
&\qquad\qquad (D^\alpha_X (\mu_1,\mu_2) \leq \rho \Rightarrow \DP \varepsilon X (\mu_1,\mu_2) \leq \delta ).
\end{align*}
In the formal proofs of conversion laws, we use this fact inside the proofs.
In fact the converse ($\impliedby$) also holds, but due to the type system of Isablle/HOL, it is hard to formalize, because of the universal quantifier $\forall X$ running over all measurable spaces.

Proof of Lemmas \ref{convRDPtoDP:Mironov} and \ref{convRDPtoDP:Balle} are done by taking lines near the boundary of $R(\alpha,\rho)$ to show the inclusions $R(\alpha,\rho) \subseteq R(\varepsilon^\ast,\delta)$ ($\varepsilon^\ast$ is determined by conversions).
They are rather straightforward interpretations of existing proofs.

The proof of Lemma \ref{convRDPtoDP:Asoodeh} given in the original paper \cite{ALCKS2021} is rather complex, but we found that we can rewrite it to proofs using privacy regions, while the authors mentioned that their approach is completely different from Balle et al.' s~\cite {DBLP:journals/corr/abs-1905-09982}.
Their approach can be understood as follows:
instead of showing $R(\alpha,\rho) \subseteq R(\varepsilon^\ast,\delta)$ directly, they prove the inclusion:
\[
[0,1] \times [0,1] \setminus R(\varepsilon^\ast,\delta) \subseteq [0,1] \times [0,1] \setminus R(\alpha,\rho).
\] 
They start the proof from the boundary of $R(\varepsilon^\ast,\delta)$ (just lines) and enable finer evaluations.
Our formal proof is based on this fact, which we borrow evaluations in the original proof.

\subsubsection{Moment Accoutants}
We here formalize Lemma \ref{lem:RDP:moment1} where we also formalize Lemma \ref{lem:RDP:moment2} (explicit verision).
\begin{isabelle}
\isakeywordONE{lemma}\isamarkupfalse%
\ Renyi{\isacharunderscore}{\kern0pt}differential{\isacharunderscore}{\kern0pt}privacy{\isacharunderscore}{\kern0pt}moments{\isacharcolon}{\kern0pt}\isanewline
\ \ \isakeywordTWO{assumes}\ {\isachardoublequoteopen}{\isadigit{0}}\ {\isasymle}\ K{\isachardoublequoteclose}\isanewline
\ \ \ \ \isakeywordTWO{and}\ {\isachardoublequoteopen}K\ {\isasymle}\ {\isasymepsilon}\ {\isacharminus}{\kern0pt}\ {\isadigit{2}}\ {\isacharasterisk}{\kern0pt}\ ln\ {\isasymdelta}\ {\isacharminus}{\kern0pt}\ {\isadigit{2}}\ {\isacharasterisk}{\kern0pt}\ sqrt{\isacharparenleft}{\kern0pt}\ {\isacharparenleft}{\kern0pt}{\isasymepsilon}\ {\isacharminus}{\kern0pt}\ ln\ {\isasymdelta}{\isacharparenright}{\kern0pt}{\isacharasterisk}{\kern0pt}\ ln{\isacharparenleft}{\kern0pt}{\isadigit{1}}{\isacharslash}{\kern0pt}{\isasymdelta}{\isacharparenright}{\kern0pt}\ {\isacharparenright}{\kern0pt}\ {\isachardoublequoteclose}\isanewline
\ \ \ \ \isakeywordTWO{and}\ {\isachardoublequoteopen}{\isadigit{0}}\ {\isacharless}{\kern0pt}\ {\isasymdelta}{\isachardoublequoteclose}\ {\isachardoublequoteopen}{\isasymdelta}\ {\isacharless}{\kern0pt}\ {\isadigit{1}}{\isachardoublequoteclose}\ \isakeywordTWO{and}\ {\isachardoublequoteopen}{\isadigit{0}}\ {\isacharless}{\kern0pt}\ {\isasymepsilon}{\isachardoublequoteclose}\isanewline
\ \ \ \ \isakeywordTWO{and}\ adj{\isacharcolon}{\kern0pt}\ {\isachardoublequoteopen}adj\ {\isasymsubseteq}\ {\isacharparenleft}{\kern0pt}space\ X{\isacharparenright}{\kern0pt}\ {\isasymtimes}\ {\isacharparenleft}{\kern0pt}space\ X{\isacharparenright}{\kern0pt}{\isachardoublequoteclose}\isanewline
\ \ \ \ \isakeywordTWO{and}\ RDPM{\isacharcolon}{\kern0pt}\ {\isachardoublequoteopen}{\isasymforall}{\isasymalpha}\ {\isachargreater}{\kern0pt}\ {\isadigit{1}}{\isachardot}{\kern0pt}\ Renyi{\isacharunderscore}{\kern0pt}differential{\isacharunderscore}{\kern0pt}privacy\ M\ adj\ {\isasymalpha}\ {\isacharparenleft}{\kern0pt}{\isasymalpha}\ {\isacharasterisk}{\kern0pt}\ K{\isacharparenright}{\kern0pt}{\isachardoublequoteclose}\isanewline
\ \ \ \ \isakeywordTWO{and}\ measM{\isacharbrackleft}{\kern0pt}measurable{\isacharbrackright}{\kern0pt}{\isacharcolon}{\kern0pt}\ {\isachardoublequoteopen}M\ {\isasymin}\ X\ {\isasymrightarrow}\isactrlsub M\ {\isacharparenleft}{\kern0pt}prob{\isacharunderscore}{\kern0pt}algebra\ Y{\isacharparenright}{\kern0pt}{\isachardoublequoteclose}\isanewline
\ \ \isakeywordTWO{shows}\ {\isachardoublequoteopen}differential{\isacharunderscore}{\kern0pt}privacy\ M\ adj\ {\isasymepsilon}\ {\isasymdelta}{\isachardoublequoteclose}
\end{isabelle}

\section{Rough Evaluation of Gaussian SGD}
In this section, we formalize the following algorithm, known as the stochastic gradient descent (SGD) mechanism with Gaussian noise.
Here is a typical formulation of such Gaussian SGD with subsampling (cf. \cite[Alg. 1]{AbadiGaussMomentAccountant} and \cite[Alg. 1]{ALCKS2021}):
\begin{algorithm}[H]
\caption{Gaussian SGD ($\nabla_\theta l,\theta,\eta,\sigma^2,C,D,L,N,T$)}\label{alg:SGD}
\begin{algorithmic}[1]
   \For{$k \gets \{1,\ldots,T\}$}
   \State $D_L \gets \mathrm{PoissonSampling}(D,L/N)$
   \For{$x_i \gets D_L$}
      \State $g \gets \nabla_\theta l (\theta,x_i)$
      \State $g' \gets \frac{C}{\| g \|} \cdot g$
      \State $z \gets \mathcal{N}( g', \sigma^2 I)$
      \State $\theta \gets \theta - \eta \cdot 1/L \cdot z$
   \EndFor
\EndFor
\State \Return $\theta$
\end{algorithmic}
\end{algorithm}

Here,  $\nabla_\theta l$ is gradient of loss function $l(\theta,-)$; $\theta$ is an initial parameter;
$0 < \eta$ is learning rate; $\sigma^2$ is variance of Gaussian noise; $0 < C$ is length-limit of gradient; $D \in (\Real^m)^\ast$ is a dataset; $N \in \Nat$ is the size of data; $L \in \Nat$ is expected batch size for subsampling; $T \in \Nat$ is a number of loops.
The subroutine $\mathrm{PoissonSampling}(D,L/N)$ samples each entry $x_i$ in the dataset $D$ independently with probability $L/N$.
We remark that $z \gets \mathcal{N}( g', \sigma^2 I)$ is a random sampling. 

\subsection{An issue on formalizing tight bounds}
This paper gives only a rough evaluation of this mechanism.
Our privacy evaluation is probably not tight.
We have not formalized a well-known tight evaluation.
We try to explain an obstacle to the formalization of a tight privacy guarantee.

The tight evaluations of Gaussian SGD in  many prior studies 
are based on the following lemma. 
\begin{lemma}[{\cite[Sec. B, Lem. 3]{AbadiGaussMomentAccountant}}]
\label{lem:RDPGaussianSubsamplings}
Let $f \colon X \to \Real^d$ with $\|f(x)\|_2 \leq 1$,
and $J$ be a sample from $[n] = \{0,\ldots,n-1\}$ where each $i \in [n]$
is chosen independently with probability $q$. 
Suppose $1 \leq \sigma$, $q < 1/16\sigma$, $\alpha \leq \sigma^2 \log (1/q\sigma)$, and that $1 < \alpha$ is integer.
Then the mechanism 
\[ 
\lambda D. \{
z_{1} \leftarrow \mathcal{N}(0,\sigma^2);\cdots; z_{|J|} \leftarrow \mathcal{N}(0,\sigma^2);
{\textstyle\sum}_{k = 1}^{|J|} z_k + f(D_{i_k})
\}
\]
is $(\alpha,q^2\alpha/(1-q)\sigma^2+ O(q^3))$-RDP.
\end{lemma}
We could not formalize this lemma in this form yet.
Moreover, for small datasets, we can give a counterexample in a corner case.

Suppose $d = 1$ and $n = 1$, and adjacency is determined by Hamming distance.
Then adjacent datasets $D_1$ and $D_2$ will be singleton lists $[x_1]$ and $[x_2]$
with $x_1 \neq x_2$.
Then the results of this mechanism for the datasets $D_k$ ($k = 1, 2$) are the following convex sum of probability measures.
\[
M(D_k) = (1-q)\mathbf{d}_0 + q\mathcal{N}(f(x_k),\sigma^2).
\]
Here, $\mathbf{d}_0$ is Dirac distribution centered at $0$ on $\Real$. 
We also assume $| f(x_1) - f(x_2) | = 1$.
Then we evaluate the R\'{e}nyi divergence between them:
\begin{align*}
&D^\alpha
(
(1-q)\mathbf{d}_0 + q\mathcal{N}(f(x_1),\sigma^2),
(1-q)\mathbf{d}_0 + q\mathcal{N}(f(x_2),\sigma^2)
)\\
&= \frac{1}{\alpha - 1} \log\left(
(1-q) + q \int_{\Real \setminus \{0\}}  
\frac{f_{\mathcal{N}(f(x_1),\sigma^2)}(y)^\alpha}{f_{\mathcal{N}(f(x_2),\sigma^2)}(y)^{\alpha-1}}dy
\right) \\
&= \frac{1}{\alpha - 1} \log\left(
(1-q) + q \exp\left((\alpha -1 )\frac{\alpha}{\sigma^2}\right) 
\right) \\
&\geq  q \frac{\alpha}{\sigma^2}.
\end{align*}
Here $f_{\mathcal{N}(f(x_k),\sigma^2)}$ ($k = 1,2$) is the density of $\mathcal{N}(f(x_1),\sigma^2)$.
The equations are obtained as follows:
We take the density functions of $M(D_k)$  ($k = 1,2$) with respect to the sum of 
$\mathbf{d}_0$ and the Borel measure, and unfold the definition of $D^\alpha$.
By splitting the integral in the definition of $D^\alpha$ with $\{0\}$ and $\Real \setminus \{0\}$, we conclude the first equation.
The second equation is obtained in the same way as in Lemma \ref{lem:Renyi:Gaussian}.

Then, the inequality is obtained by a simple calculation.
Since $q$ is quite smaller than $1$, the resulting value  $\alpha q/\sigma^2$ is greater than $q^2\alpha/(1-q)\sigma^2+ O(q^3)$.
We have this issue regardless of assumptions 
$1 \leq \sigma$, $q < 1/16\sigma$, $\alpha \leq \sigma^2 \log (1/q\sigma)$, $1 < \alpha$ : integer. 

For another example, 
Suppose $d = 1$ and $n = 1$, and that the adjacency is determined by insertion  $D_2 = D_1 \cup \{x\}$.
Then, we consider $D_1 = \{\}$ and $D_2 = \{x\}$.
In similar way as above, we can calculate the resulting R\'{e}nyi divergences: 
\begin{align*}
&D^\alpha(\mathbf{d}_0,(1-q)\mathbf{d}_0 + q\mathcal{N}(f(x),\sigma^2)) \\
&= \frac{1}{\alpha - 1} \log \left( (1-q)^{1-\alpha} + q \int_{\Real \setminus \{0\}} \frac{0}{f_{\mathcal{N}(f(x_1),\sigma^2)}(y)^{\alpha-1}} dy \right) \\
&= - \log (1-q)\\
&D^\alpha((1-q)\mathbf{d}_0 + q\mathcal{N}(f(x),\sigma^2),\mathbf{d}_0) \\
&= \frac{1}{\alpha - 1} \log \left( (1-q)^{\alpha} + q \int_{\Real \setminus \{0\}} \frac{f_{\mathcal{N}(f(x_1),\sigma^2)}(y)^{\alpha}}{0^{\alpha - 1}} dy \right) \\
&= \frac{1}{\alpha - 1} \log \left( (1-q)^{\alpha} + q \infty \right) = \infty
\end{align*}
Therefore, for both definitions of adjacency, 
$(\alpha,q^2\alpha/(1-q)\sigma^2+ O(q^3))$-RDP fails when the datasets are too small, and the initial value is fixed.

The original proof of Lemma \ref{lem:RDPGaussianSubsamplings} is based on showing the following inequality:
\begin{align*}
&D^\alpha(\mathcal{N}(0,\sigma^2),(1-q)\mathcal{N}(0,\sigma^2) + q\mathcal{N}(1,\sigma^2))\\
&\leq q^2\alpha/(1-q)\sigma^2+ O(q^3)).
\end{align*}
We must assume a mixture distribution of two Gaussian distributions with the same variance $\sigma^2$,
to cancel the terms of order $ O(q^1))$ in the evaluation of the divergence.
However, showing the above inequality does not connect well to the evaluation of the entire mechanism,
even if we assume that the initial value is also sampled from a Gaussian distribution.

Intuitively, for the insertion adjacency $D_2 = D_1 \cup \{x\}$, the distribution 
$\mathcal{N}(0,\sigma^2)$ in the above inequality indicates the branch that the differing element $x$ 
was not sampled in $J$, and  $\mathcal{N}(1,\sigma^2)$ indicates the one that $x$ was sampled.
In the second branch, from the element $x$, a new random sampling $z_x \leftarrow \mathcal{N}(x,\sigma_*^2)$
must be added to the total, other than $z_x$. Hence, the variance of the second branch must be greater than the first one by $\sigma_*^2$.
Hence, what we need to evaluate is
\[
D^\alpha(\mathcal{N}(0,\sigma_1^2),(1-q)\mathcal{N}(0,\sigma_1^2) + q\mathcal{N}(1,\sigma_2^2))
\]
where $\sigma_1^2$ and $\sigma_2^2$ are different (moreover $\sigma_1^2 < \sigma_2^2$).
We postpone the evaluation of the R\'{e}nyi divergence between such mixture distributions to future work.

\subsection{A rule for subsampled RDP}
Privacy amplification by subsampling~\cite{3327345.3327525,pmlr-v89-wang19b,AbadiGaussMomentAccountant} is an important technique for privacy-preserving mechanisms of large populations.
In prior studies, these evaluations are rather complicated, hence formalizing them requires much more work.
In this paper, we formalize a simpler (but rough) rule for RDP based on the joint-convexity of R\'{e}nyi divergence instead.
From the joint-convexity of $f$-divergences,
\begin{lemma}
\label{lem:convexRDP2}
For $\mu,\nu_1,\nu_2 \in \prob (X)$, $1 < \alpha$ and $0 < q  < 1$, 
\begin{align*}
&D^\alpha((1-q)\mu + q\nu_1,(1-q)\mu + q\nu_2) \\
&\leq \smash{\frac{1}{\alpha - 1}} \log 
\left(
(1-q)+ q \exp((\alpha-1) D^\alpha(\mu_2,\nu_2)
\right).
\end{align*}
\end{lemma}
It gives an evaluation of the amplification of RDP by subsampling as follows:
Intuitively, subsampling deletes differing entries $x_1\neq x_2$ in adjacent datasets $D_1$ and $D_2$ with probability $q$.
Then the result can be formulated as $(1-q)\mu + q\nu_k$ where $\mu$ and $\nu_k$ are the results where $x_k$ is deleted and remains, respectively ($k=1,2$).

We can determine the variance $\sigma^2$ of Gaussian noise making the Gaussian SGD with Poisson subsampling $(\varepsilon,\delta)$-DP, by combineing Lemmas \ref{lem:convexRDP2} and  \ref{lem:RDP:moment2} and the following lemma.
\begin{lemma}
\label{lem:linearapproximation:convexRDP}
For $f(r) = \frac{1}{\alpha - 1} \log \left((1-q)+ q \exp((\alpha-1) r \right)$,
\begin{align*}
r \leq \frac{\log(1/\sqrt{q})}{\alpha - 1} & \implies f(r) \leq \frac{\log(1-q+\sqrt{q})}{\log(1/\sqrt{q})}r .
\end{align*}
\end{lemma}
This lemma can be proved from the convexity of $f(r)$.

These rules can be mechanized in Isabelle/HOL.
For example, we formalize Lemma \ref{lem:convexRDP2} as follows:
\begin{isabelle}
\isakeywordONE{lemma}\isamarkupfalse%
\ {\isacharparenleft}{\kern0pt}\isakeywordTWO{in}\ comparable{\isacharunderscore}{\kern0pt}finite{\isacharunderscore}{\kern0pt}measures{\isacharparenright}{\kern0pt}\ Renyi{\isacharunderscore}{\kern0pt}div{\isacharunderscore}{\kern0pt}convex{\isacharunderscore}{\kern0pt}sum{\isadigit{2}}{\isacharcolon}{\kern0pt}\isanewline
\isakeywordTWO{assumes}\ a{\isacharcolon}{\kern0pt}\ {\isachardoublequoteopen}{\isadigit{1}}\ {\isacharless}{\kern0pt}\ {\isacharparenleft}{\kern0pt}{\isasymalpha}{\isacharcolon}{\kern0pt}{\isacharcolon}{\kern0pt}real{\isacharparenright}{\kern0pt}{\isachardoublequoteclose}
\ \isakeywordTWO{and}\ b{\isacharprime}{\kern0pt}{\isacharcolon}{\kern0pt}\ {\isachardoublequoteopen}{\isadigit{0}}\ {\isasymle}\ {\isacharparenleft}{\kern0pt}b{\isacharcolon}{\kern0pt}{\isacharcolon}{\kern0pt}real{\isacharparenright}{\kern0pt}{\isachardoublequoteclose}\ {\isachardoublequoteopen}b\ {\isasymle}\ {\isadigit{1}}{\isachardoublequoteclose}\isanewline
\ \ \isakeywordTWO{and}\ M{\isadigit{2}}{\isacharcolon}{\kern0pt}\ {\isachardoublequoteopen}M{\isadigit{2}}\ {\isasymin}\ space{\isacharparenleft}{\kern0pt}prob{\isacharunderscore}{\kern0pt}algebra\ X{\isacharparenright}{\kern0pt}{\isachardoublequoteclose}\ \isakeywordTWO{and}\ {\isachardoublequoteopen}\ {\isacharparenleft}{\kern0pt}Renyi{\isacharunderscore}{\kern0pt}div\ {\isasymalpha}\ \ M\ N{\isacharparenright}{\kern0pt}\ {\isasymle}\ R{\isachardoublequoteclose}\isanewline
\isakeywordTWO{shows}\ {\isachardoublequoteopen}{\isacharparenleft}{\kern0pt}Renyi{\isacharunderscore}{\kern0pt}div\ {\isasymalpha}\ {\isacharparenleft}{\kern0pt}conv{\isacharunderscore}{\kern0pt}measure\ b\ M\ M{\isadigit{2}}{\isacharparenright}{\kern0pt}\ {\isacharparenleft}{\kern0pt}conv{\isacharunderscore}{\kern0pt}measure\ b\ N\ M{\isadigit{2}}{\isacharparenright}{\kern0pt}{\isacharparenright}{\kern0pt}\ {\isasymle}\ Renyi{\isacharunderscore}{\kern0pt}log{\isacharunderscore}{\kern0pt}part\ {\isasymalpha}\ {\isacharparenleft}{\kern0pt}ennreal\ b\ {\isacharasterisk}{\kern0pt}\ inv{\isacharunderscore}{\kern0pt}Renyi{\isacharunderscore}{\kern0pt}log{\isacharunderscore}{\kern0pt}part\ {\isasymalpha}\ R\ {\isacharplus}{\kern0pt}\ \ ennreal\ {\isacharparenleft}{\kern0pt}{\isadigit{1}}\ {\isacharminus}{\kern0pt}\ b{\isacharparenright}{\kern0pt}{\isacharparenright}{\kern0pt}{\isachardoublequoteclose}
\end{isabelle}
%
\subsection{A sketch of formalization}
Let's move on to our formalization. 
In our formalization, the ratio $L/N$ in Algorithm \ref{alg:SGD} is abstracted as $q$, and $\eta \cdot 1/L$ is simplified as $\eta$. 
For simplicity we fix \isa{\isasymtheta}, since its distribution is not given in prior studies.
We fix \isa{{\isachardoublequoteopen}{\isacharparenleft}{\kern0pt}{\isasymlambda}{\isacharparenleft}{\kern0pt}{\isasymtheta}{\isacharcomma}{\kern0pt}\ x{\isacharparenright}{\kern0pt}{\isachardot}{\kern0pt}\ grad\ {\isasymtheta}\ x{\isacharparenright}{\kern0pt}\ {\isasymin}\ borel\ {\isasymOtimes}\isactrlsub M\ borel\ {\isasymrightarrow}\isactrlsub M\ borel{\isachardoublequoteclose}} for formalization of $\nabla_\theta l$.

\subsubsection{Implementation of programs}
We consider datasets consisting of \isa{m}-dimensional tuples.
We formalize the space $\Real^m$ of  \isa{m}-dimensional tuples as the following \isa{Sp}.
\begin{isabelle}
\isakeywordONE{abbreviation}\isamarkupfalse%
\ {\isachardoublequoteopen}sM\ {\isasymequiv}\ {\isacharbraceleft}{\kern0pt}{\isadigit{0}}{\isachardot}{\kern0pt}{\isachardot}{\kern0pt}m{\isacharminus}{\kern0pt}{\isadigit{1}}{\isacharbraceright}{\kern0pt}{\isachardoublequoteclose}
\end{isabelle}
\begin{isabelle}
\isakeywordONE{abbreviation}\isamarkupfalse%
\ {\isachardoublequoteopen}Sp\ {\isasymequiv}\ {\isacharparenleft}{\kern0pt}{\isasymPi}\isactrlsub M\ x{\isasymin}sM{\isachardot}{\kern0pt}\ {\isacharparenleft}{\kern0pt}borel{\isacharcolon}{\kern0pt}{\isacharcolon}{\kern0pt}real\ measure{\isacharparenright}{\kern0pt}{\isacharparenright}{\kern0pt}{\isachardoublequoteclose}
\end{isabelle}
We then implement Algorithm \ref{alg:SGD}.

We first implement Lines 4--7 of Algorithm \ref{alg:SGD} (a loop step).
\begin{isabelle}
\isakeywordONE{definition}\isamarkupfalse%
\ SGD{\isacharunderscore}{\kern0pt}step\ \isakeywordTWO{where}\isanewline
\ \ {\isachardoublequoteopen}\ SGD{\isacharunderscore}{\kern0pt}step\ v{\isasymtheta}\ vx\ {\isacharequal}{\kern0pt}\ do\ {\isacharbraceleft}{\kern0pt}\isanewline
\ \ \ \ \ \ let\ vgx\ {\isacharequal}{\kern0pt}\ grad\ v{\isasymtheta}\ vx{\isacharsemicolon}{\kern0pt}\ \isanewline
\ \ \ \ \ \ let\ vgx{\isadigit{2}}\ {\isacharequal}{\kern0pt}\ {\isacharparenleft}{\kern0pt}if\ Sp{\isacharunderscore}{\kern0pt}L{\isadigit{2}}{\isacharunderscore}{\kern0pt}set\ vgx\ {\isachargreater}{\kern0pt}\ C\ then\ {\isacharparenleft}{\kern0pt}{\isasymlambda}k{\isasymin}sM{\isachardot}{\kern0pt}\ {\isacharparenleft}{\kern0pt}C{\isacharslash}{\kern0pt}\ Sp{\isacharunderscore}{\kern0pt}L{\isadigit{2}}{\isacharunderscore}{\kern0pt}set\ vgx{\isacharparenright}{\kern0pt}\ {\isacharasterisk}{\kern0pt}\ vgx\ k{\isacharparenright}{\kern0pt}\ else\ vgx{\isacharparenright}{\kern0pt}{\isacharsemicolon}{\kern0pt}\ \isanewline
\ \ \ \ \ \ vy\ {\isasymleftarrow}\ Gaussian{\isacharunderscore}{\kern0pt}vec\ {\isasymsigma}\ m\ vgx{\isadigit{2}}{\isacharsemicolon}{\kern0pt}\ \isanewline
\ \ \ \ \ \ let\ v{\isasymtheta}new\ {\isacharequal}{\kern0pt}\ {\isacharparenleft}{\kern0pt}{\isasymlambda}k{\isasymin}sM{\isachardot}{\kern0pt}\ v{\isasymtheta}\ k\ {\isacharminus}{\kern0pt}\ rate\ {\isacharasterisk}{\kern0pt}\ vy\ k{\isacharparenright}{\kern0pt}{\isacharsemicolon}{\kern0pt}\ \isanewline
\ \ \ \ \ \ return\ Sp\ v{\isasymtheta}new\isanewline
\ \ {\isacharbraceright}{\kern0pt}{\isachardoublequoteclose}
\end{isabelle}

Next, we implement Lines 3--8 of Algorithm \ref{alg:SGD}(batch loop).
\begin{isabelle}
\isakeywordONE{primrec}\isamarkupfalse%
\ SGD{\isacharunderscore}{\kern0pt}batch\ \isakeywordTWO{where}\isanewline
\ \ {\isachardoublequoteopen}SGD{\isacharunderscore}{\kern0pt}batch\ {\isasymtheta}\ {\isacharbrackleft}{\kern0pt}{\isacharbrackright}{\kern0pt}\ {\isacharequal}{\kern0pt}\ {\isacharparenleft}{\kern0pt}return\ Sp\ {\isasymtheta}{\isacharparenright}{\kern0pt}{\isachardoublequoteclose}\isanewline
{\isacharbar}{\kern0pt}\ {\isachardoublequoteopen}SGD{\isacharunderscore}{\kern0pt}batch\ {\isasymtheta}\ {\isacharparenleft}{\kern0pt}x\ {\isacharhash}{\kern0pt}\ xs{\isacharparenright}{\kern0pt}\ {\isacharequal}{\kern0pt}\ {\isacharparenleft}{\kern0pt}SGD{\isacharunderscore}{\kern0pt}step\ {\isasymtheta}\ x{\isacharparenright}{\kern0pt}\ {\isasymbind}\ {\isacharparenleft}{\kern0pt}{\isasymlambda}{\isasymtheta}{\isachardot}{\kern0pt}\ SGD{\isacharunderscore}{\kern0pt}batch\ {\isasymtheta}\ xs{\isacharparenright}{\kern0pt}{\isachardoublequoteclose}
\end{isabelle}

Next, we implement the subroutine $\mathrm{PoissonSampling}(D,b)$ (Line 2 of Algorithm \ref{alg:SGD}).
We assume \isa{0 < b} and \isa{b < 1}.
\begin{isabelle}
\isakeywordONE{primrec}\isamarkupfalse%
\ Poisson{\isacharunderscore}{\kern0pt}Sample\ \isakeywordTWO{where}\isanewline
\ \ {\isachardoublequoteopen}Poisson{\isacharunderscore}{\kern0pt}Sample\ {\isacharbrackleft}{\kern0pt}{\isacharbrackright}{\kern0pt}\ {\isacharequal}{\kern0pt}\ return\ {\isacharparenleft}{\kern0pt}listM\ M{\isacharparenright}{\kern0pt}\ {\isacharbrackleft}{\kern0pt}{\isacharbrackright}{\kern0pt}{\isachardoublequoteclose}\isanewline
{\isacharbar}{\kern0pt}\ {\isachardoublequoteopen}Poisson{\isacharunderscore}{\kern0pt}Sample\ {\isacharparenleft}{\kern0pt}x{\isacharhash}{\kern0pt}xs{\isacharparenright}{\kern0pt}\ {\isacharequal}{\kern0pt}\ do{\isacharbraceleft}{\kern0pt}xs{\isadigit{2}}\ {\isasymleftarrow}\ \ Poisson{\isacharunderscore}{\kern0pt}Sample\ xs{\isacharsemicolon}{\kern0pt}\ c\ {\isasymleftarrow}\ bern{\isacharunderscore}{\kern0pt}dist\ {\isacharsemicolon}{\kern0pt}\ return\ {\isacharparenleft}{\kern0pt}listM\ M{\isacharparenright}{\kern0pt}\ {\isacharparenleft}{\kern0pt}if\ c\ then\ x\ {\isacharhash}{\kern0pt}\ xs{\isadigit{2}}\ else\ xs{\isadigit{2}}{\isacharparenright}{\kern0pt}{\isacharbraceright}{\kern0pt}{\isachardoublequoteclose}
\end{isabelle}
\begin{isabelle}
\isakeywordONE{abbreviation}\isamarkupfalse%
\ {\isachardoublequoteopen}bern{\isacharunderscore}{\kern0pt}dist\ {\isasymequiv}\ measure{\isacharunderscore}{\kern0pt}pmf\ {\isacharparenleft}{\kern0pt}bernoulli{\isacharunderscore}{\kern0pt}pmf\ b{\isacharparenright}{\kern0pt}{\isachardoublequoteclose}
\end{isabelle}
Here, \isa{M} is a measurable space for the domain of data points.
A dataset is formalized as  \isa{space (listM\ M)}.
The parameter \isa{b} is the probability of sampling of each data point, and \isa{bern{\isacharunderscore}{\kern0pt}dist} is just the coin flip with probabilty \isa{b}.
We here remind another formulation of convex combinations by coin flips. 
\begin{isabelle}
\isakeywordONE{lemma}\isamarkupfalse%
\ bern{\isacharunderscore}{\kern0pt}dist{\isacharunderscore}{\kern0pt}conv{\isacharunderscore}{\kern0pt}measure{\isacharcolon}{\kern0pt}\isanewline
\ \ \isakeywordTWO{assumes}\ {\isachardoublequoteopen}M\ {\isasymin}\ space\ {\isacharparenleft}{\kern0pt}prob{\isacharunderscore}{\kern0pt}algebra\ L{\isacharparenright}{\kern0pt}{\isachardoublequoteclose}\isanewline
\ \ \ \ \isakeywordTWO{and}\ {\isachardoublequoteopen}N\ {\isasymin}\ space\ {\isacharparenleft}{\kern0pt}prob{\isacharunderscore}{\kern0pt}algebra\ L{\isacharparenright}{\kern0pt}{\isachardoublequoteclose}\isanewline
\ \ \isakeywordTWO{shows}\ {\isachardoublequoteopen}do{\isacharbraceleft}{\kern0pt}\ c\ {\isasymleftarrow}\ bern{\isacharunderscore}{\kern0pt}dist\ {\isacharsemicolon}{\kern0pt}\ if\ c\ then\ M\ else\ N{\isacharbraceright}{\kern0pt}\isanewline
\ \ \ \ {\isacharequal}{\kern0pt}\ conv{\isacharunderscore}{\kern0pt}measure\ b\ M\ N{\isachardoublequoteclose}
\end{isabelle}

We then implement Lines 3--8 of Algorithm \ref{alg:SGD}.
\begin{isabelle}
\isakeywordONE{definition}\isamarkupfalse%
\ SGD{\isacharunderscore}{\kern0pt}Poisson\ \isakeywordTWO{where}\isanewline
\ {\isachardoublequoteopen}SGD{\isacharunderscore}{\kern0pt}Poisson\ {\isasymtheta}\ D\ {\isacharequal}{\kern0pt}\ do{\isacharbraceleft}{\kern0pt}xs\ {\isasymleftarrow}\ {\isacharparenleft}{\kern0pt}Poisson{\isacharunderscore}{\kern0pt}Sample\ D{\isacharparenright}{\kern0pt}{\isacharsemicolon}{\kern0pt}\ SGD{\isacharunderscore}{\kern0pt}batch\ {\isasymtheta}\ xs{\isacharbraceright}{\kern0pt}{\isachardoublequoteclose}
\end{isabelle}

Finally, we implement the entire mechanism by
\begin{isabelle}
\isakeywordONE{primrec}\isamarkupfalse%
\ SGD{\isacharunderscore}{\kern0pt}Poisson{\isacharunderscore}{\kern0pt}iter\ \isakeywordTWO{where}\isanewline
\ \ {\isachardoublequoteopen}\ SGD{\isacharunderscore}{\kern0pt}Poisson{\isacharunderscore}{\kern0pt}iter\ {\isasymtheta}\ D\ {\isadigit{0}}\ {\isacharequal}{\kern0pt}\ {\isacharparenleft}{\kern0pt}return\ borel\ {\isasymtheta}{\isacharparenright}{\kern0pt}{\isachardoublequoteclose}\isanewline
{\isacharbar}{\kern0pt}\ {\isachardoublequoteopen}SGD{\isacharunderscore}{\kern0pt}Poisson{\isacharunderscore}{\kern0pt}iter\ {\isasymtheta}\ D\ {\isacharparenleft}{\kern0pt}Suc\ k{\isacharparenright}{\kern0pt}\isanewline
\ \ \ {\isacharequal}{\kern0pt}\ do{\isacharbraceleft}{\kern0pt}{\isasymtheta}{\isacharprime}{\kern0pt}\ {\isasymleftarrow}\ SGD{\isacharunderscore}{\kern0pt}Poisson\ {\isasymtheta}\ D\ {\isacharsemicolon}{\kern0pt}\ {\isacharparenleft}{\kern0pt}SGD{\isacharunderscore}{\kern0pt}Poisson{\isacharunderscore}{\kern0pt}iter\ {\isasymtheta}{\isacharprime}{\kern0pt}\ D\ k{\isacharparenright}{\kern0pt}{\isacharbraceright}{\kern0pt}{\isachardoublequoteclose}
\end{isabelle}

\subsubsection{Evaluations for DP and RDP}
We consider the adjacency with Hamming distance.
We formalize a rough evaluation for RDP of
\isa{ SGD{\isacharunderscore}{\kern0pt}Poisson{\isacharunderscore}{\kern0pt}iter}.
Below, \isa{HLists} is the set of \isa{Sp}-lists of length \isa{N} and \isa{Hdist} is the Hamming distance.
\begin{isabelle}
\isakeywordONE{lemma}\isamarkupfalse%
\ RDP{\isacharunderscore}{\kern0pt}SGD{\isacharunderscore}{\kern0pt}Poisson{\isacharunderscore}{\kern0pt}iter{\isacharcolon}{\kern0pt}\isanewline
\ \ \isakeywordTWO{assumes}\ a{\isacharcolon}{\kern0pt}\ {\isachardoublequoteopen}{\isadigit{1}}\ {\isacharless}{\kern0pt}\ {\isasymalpha}{\isachardoublequoteclose}
\ \isakeywordTWO{and}\ {\isachardoublequoteopen}{\isasymtheta}\ {\isasymin}\ space\ Sp{\isachardoublequoteclose}\isanewline
\ \ \isakeywordTWO{defines}\ {\isachardoublequoteopen}\ \ adj\ {\isasymequiv}\ {\isacharbraceleft}{\kern0pt}{\isacharparenleft}{\kern0pt}xs{\isacharcomma}{\kern0pt}ys{\isacharparenright}{\kern0pt}{\isacharbar}{\kern0pt}\ xs\ ys{\isachardot}{\kern0pt}\ xs\ {\isasymin}\ HLists\ {\isasymand}\ ys\ {\isasymin}\ HLists\ {\isasymand}\ Hdist\ xs\ ys\ {\isacharequal}{\kern0pt}\ {\isadigit{1}}{\isacharbraceright}{\kern0pt}{\isachardoublequoteclose}\isanewline
\ \ \isakeywordTWO{shows}\ {\isachardoublequoteopen}Renyi{\isacharunderscore}{\kern0pt}differential{\isacharunderscore}{\kern0pt}privacy\ {\isacharparenleft}{\kern0pt}{\isasymlambda}xs{\isachardot}{\kern0pt}\ SGD{\isacharunderscore}{\kern0pt}Poisson{\isacharunderscore}{\kern0pt}iter\ {\isasymtheta}\ xs\ T{\isacharparenright}{\kern0pt}\ \isanewline
\ \ \ \ adj\ {\isasymalpha}\ {\isacharparenleft}{\kern0pt}T\ {\isacharasterisk}{\kern0pt}\ {\isacharparenleft}{\kern0pt}ln\ {\isacharparenleft}{\kern0pt}b\ {\isacharasterisk}{\kern0pt}\ exp{\isacharparenleft}{\kern0pt}{\isacharparenleft}{\kern0pt}{\isadigit{2}}\ {\isacharasterisk}{\kern0pt}\ {\isacharparenleft}{\kern0pt}{\isasymalpha}\ {\isacharasterisk}{\kern0pt}\ C\isactrlsup {\isadigit{2}}{\isacharparenright}{\kern0pt}\ {\isacharslash}{\kern0pt}\ {\isasymsigma}\isactrlsup {\isadigit{2}}{\isacharparenright}{\kern0pt}\ {\isacharasterisk}{\kern0pt}\ {\isacharparenleft}{\kern0pt}{\isasymalpha}{\isacharminus}{\kern0pt}{\isadigit{1}}{\isacharparenright}{\kern0pt}{\isacharparenright}{\kern0pt}\ {\isacharplus}{\kern0pt}\ {\isacharparenleft}{\kern0pt}{\isadigit{1}}\ {\isacharminus}{\kern0pt}\ b{\isacharparenright}{\kern0pt}{\isacharparenright}{\kern0pt}\ {\isacharslash}{\kern0pt}\ {\isacharparenleft}{\kern0pt}{\isasymalpha}{\isacharminus}{\kern0pt}{\isadigit{1}}{\isacharparenright}{\kern0pt}{\isacharparenright}{\kern0pt}{\isacharparenright}{\kern0pt}{\isachardoublequoteclose}
\end{isabelle}
Combining this result and formalization of Lemmas \ref{lem:RDP:moment2} and \ref{lem:linearapproximation:convexRDP}, 
we can formalize a condition of $\sigma$, $C$ and $T$ to guarantee $(\varepsilon,\delta)$-DP.
\begin{isabelle}
\isakeywordONE{lemma}\isamarkupfalse%
\ DP{\isacharunderscore}{\kern0pt}SGD{\isacharunderscore}{\kern0pt}Poisson{\isacharunderscore}{\kern0pt}iter{\isacharcolon}{\kern0pt}\isanewline
\ \ \isakeywordTWO{assumes}\ {\isachardoublequoteopen}{\isasymtheta}\ {\isasymin}\ space\ Sp{\isachardoublequoteclose}\
  \isakeywordTWO{and}\ {\isachardoublequoteopen}{\isadigit{0}}\ {\isacharless}{\kern0pt}\ {\isasymdelta}{\isachardoublequoteclose}\ {\isachardoublequoteopen}{\isasymdelta}\ {\isacharless}{\kern0pt}\ {\isadigit{1}}{\isachardoublequoteclose}\ \isakeywordTWO{and}\ {\isachardoublequoteopen}{\isadigit{0}}\ {\isacharless}{\kern0pt}\ {\isasymepsilon}{\isachardoublequoteclose}\isanewline
\ \ \ \ \isakeywordTWO{and}\ cond{\isadigit{2}}{\isacharcolon}{\kern0pt}\ {\isachardoublequoteopen}{\isadigit{2}}\ {\isacharasterisk}{\kern0pt}\ C\isactrlsup {\isadigit{2}}\ {\isacharslash}{\kern0pt}\ {\isasymsigma}\isactrlsup {\isadigit{2}}\ {\isasymle}\ {\isacharminus}{\kern0pt}\ ln\ {\isacharparenleft}{\kern0pt}sqrt\ b{\isacharparenright}{\kern0pt}\ {\isacharslash}{\kern0pt}\ {\isacharparenleft}{\kern0pt}{\isacharparenleft}{\kern0pt}{\isadigit{1}}\ {\isacharminus}{\kern0pt}\ ln\ {\isasymdelta}\ {\isacharslash}{\kern0pt}\ {\isasymepsilon}\ {\isacharplus}{\kern0pt}\ sqrt\ {\isacharparenleft}{\kern0pt}{\isacharparenleft}{\kern0pt}{\isasymepsilon}\ {\isacharminus}{\kern0pt}\ ln\ {\isasymdelta}{\isacharparenright}{\kern0pt}\ {\isacharasterisk}{\kern0pt}\ {\isacharminus}{\kern0pt}\ ln\ {\isasymdelta}{\isacharparenright}{\kern0pt}\ {\isacharslash}{\kern0pt}\ {\isasymepsilon}\ {\isacharminus}{\kern0pt}\ {\isadigit{1}}{\isacharparenright}{\kern0pt}\ {\isacharasterisk}{\kern0pt}\ {\isacharparenleft}{\kern0pt}{\isadigit{1}}\ {\isacharminus}{\kern0pt}\ ln\ {\isasymdelta}\ {\isacharslash}{\kern0pt}\ {\isasymepsilon}\ {\isacharplus}{\kern0pt}\ sqrt\ {\isacharparenleft}{\kern0pt}{\isacharparenleft}{\kern0pt}{\isasymepsilon}\ {\isacharminus}{\kern0pt}\ ln\ {\isasymdelta}{\isacharparenright}{\kern0pt}\ {\isacharasterisk}{\kern0pt}\ {\isacharminus}{\kern0pt}\ ln\ {\isasymdelta}{\isacharparenright}{\kern0pt}\ {\isacharslash}{\kern0pt}\ {\isasymepsilon}{\isacharparenright}{\kern0pt}{\isacharparenright}{\kern0pt}{\isachardoublequoteclose}\isanewline
\ \ \ \ \isakeywordTWO{and}\ cond{\isadigit{1}}{\isacharcolon}{\kern0pt}\ {\isachardoublequoteopen}{\isacharparenleft}{\kern0pt}{\isadigit{2}}\ {\isacharasterisk}{\kern0pt}\ real\ T\ {\isacharasterisk}{\kern0pt}\ C\isactrlsup {\isadigit{2}}{\isacharparenright}{\kern0pt}\ {\isacharslash}{\kern0pt}\ {\isasymsigma}\isactrlsup {\isadigit{2}}\ {\isasymle}\ {\isacharparenleft}{\kern0pt}{\isasymepsilon}\ {\isacharminus}{\kern0pt}\ {\isadigit{2}}\ {\isacharasterisk}{\kern0pt}\ ln\ {\isasymdelta}\ {\isacharminus}{\kern0pt}\ {\isadigit{2}}\ {\isacharasterisk}{\kern0pt}\ sqrt\ {\isacharparenleft}{\kern0pt}{\isacharparenleft}{\kern0pt}{\isasymepsilon}\ {\isacharminus}{\kern0pt}\ ln\ {\isasymdelta}{\isacharparenright}{\kern0pt}\ {\isacharasterisk}{\kern0pt}\ ln\ {\isacharparenleft}{\kern0pt}{\isadigit{1}}\ {\isacharslash}{\kern0pt}\ {\isasymdelta}{\isacharparenright}{\kern0pt}{\isacharparenright}{\kern0pt}{\isacharparenright}{\kern0pt}\ {\isacharasterisk}{\kern0pt}\ \ {\isacharparenleft}{\kern0pt}\ {\isacharminus}{\kern0pt}\ ln\ {\isacharparenleft}{\kern0pt}sqrt\ b{\isacharparenright}{\kern0pt}\ {\isacharslash}{\kern0pt}\ ln\ {\isacharparenleft}{\kern0pt}{\isadigit{1}}\ {\isacharminus}{\kern0pt}\ b\ {\isacharplus}{\kern0pt}\ sqrt\ b{\isacharparenright}{\kern0pt}{\isacharparenright}{\kern0pt}{\isachardoublequoteclose}\isanewline
\ \ \isakeywordTWO{defines}\ {\isachardoublequoteopen}\ \ adj\ {\isasymequiv}\ {\isacharbraceleft}{\kern0pt}{\isacharparenleft}{\kern0pt}xs{\isacharcomma}{\kern0pt}ys{\isacharparenright}{\kern0pt}{\isacharbar}{\kern0pt}\ xs\ ys{\isachardot}{\kern0pt}\ xs\ {\isasymin}\ HLists\ {\isasymand}\ ys\ {\isasymin}\ HLists\ {\isasymand}\ Hdist\ xs\ ys\ {\isacharequal}{\kern0pt}\ {\isadigit{1}}{\isacharbraceright}{\kern0pt}{\isachardoublequoteclose}\isanewline
\ \ \isakeywordTWO{shows}\ {\isachardoublequoteopen}differential{\isacharunderscore}{\kern0pt}privacy\ {\isacharparenleft}{\kern0pt}{\isasymlambda}xs{\isachardot}{\kern0pt}\ SGD{\isacharunderscore}{\kern0pt}Poisson{\isacharunderscore}{\kern0pt}iter\ {\isasymtheta}\ xs\ T{\isacharparenright}{\kern0pt}\ adj\ {\isasymepsilon}\ {\isasymdelta}{\isachardoublequoteclose}
\end{isabelle}

\subsubsection{Sketch of formal proof}
We first evaluate RDP of a step of loop (\isa{SGD{\isacharunderscore}{\kern0pt}step}).
It is a typical Gaussian mechanism, where calculation and clipping of gradient has sensitivity $2C$.
We just evaluate the R\'{e}nyi divergence.
\begin{isabelle}
\isakeywordONE{lemma}\isamarkupfalse%
\ Renyi{\isacharunderscore}{\kern0pt}div{\isacharunderscore}{\kern0pt}SGD{\isacharunderscore}{\kern0pt}step{\isacharcolon}{\kern0pt}\isanewline
\ \ \isakeywordTWO{assumes}\ a{\isacharcolon}{\kern0pt}\ {\isachardoublequoteopen}{\isadigit{1}}\ {\isacharless}{\kern0pt}\ {\isasymalpha}{\isachardoublequoteclose}
\ \isakeywordTWO{and}\ {\isachardoublequoteopen}v{\isasymtheta}\ {\isasymin}\ space\ Sp{\isachardoublequoteclose}\ {\isachardoublequoteopen}vx{\isadigit{1}}\ {\isasymin}\ space\ Sp{\isachardoublequoteclose}\ {\isachardoublequoteopen}vx{\isadigit{2}}\ {\isasymin}\ space\ Sp{\isachardoublequoteclose}\isanewline
\ \ \isakeywordTWO{shows}\ {\isachardoublequoteopen}Renyi{\isacharunderscore}{\kern0pt}div\ {\isasymalpha}\ {\isacharparenleft}{\kern0pt}SGD{\isacharunderscore}{\kern0pt}step\ v{\isasymtheta}\ vx{\isadigit{1}}{\isacharparenright}{\kern0pt}\ {\isacharparenleft}{\kern0pt}SGD{\isacharunderscore}{\kern0pt}step\ v{\isasymtheta}\ vx{\isadigit{2}}{\isacharparenright}{\kern0pt}\ {\isasymle}\ {\isacharparenleft}{\kern0pt}{\isadigit{2}}\ {\isacharasterisk}{\kern0pt}\ {\isacharparenleft}{\kern0pt}{\isasymalpha}\ {\isacharasterisk}{\kern0pt}\ C\isactrlsup {\isadigit{2}}{\isacharparenright}{\kern0pt}\ \ {\isacharslash}{\kern0pt}\ {\isasymsigma}\isactrlsup {\isadigit{2}}{\isacharparenright}{\kern0pt}{\isachardoublequoteclose}
\end{isabelle}

We next evaluate the batch loop (\isa{SGD{\isacharunderscore}{\kern0pt}batch}) and Poisson subsampling (\isa{Poisson{\isacharunderscore}{\kern0pt}Sample}).
Both of them have similar structures.
For $D = \{x_1,\ldots,x_k\}$, each of them can be reforemulated as bind-iterations 
of some measurable function $g(x_i,-) \colon \isa{X} \to \prob(\isa{X})$ and distribution $P \in \prob(X)$:
\[
P \bind g(x_1,-) \bind \cdots \bind g(x_k,-).
\]
From the associativity of the Giry monad bind, we can formalize that each of them preserves appends, hence so does their sequential composition \isa{SGD{\isacharunderscore}{\kern0pt}Poisson}:
\begin{isabelle}
\isakeywordONE{lemma}\isamarkupfalse%
\ SGD{\isacharunderscore}{\kern0pt}Poisson{\isacharunderscore}{\kern0pt}append{\isacharcolon}{\kern0pt}\isanewline
\ \ \isakeywordTWO{assumes}\ {\isachardoublequoteopen}xs\ {\isasymin}\ space\ {\isacharparenleft}{\kern0pt}listM\ Sp{\isacharparenright}{\kern0pt}{\isachardoublequoteclose}\ {\isachardoublequoteopen}ys\ {\isasymin}\ space\ {\isacharparenleft}{\kern0pt}listM\ Sp{\isacharparenright}{\kern0pt}{\isachardoublequoteclose}\ {\isachardoublequoteopen}{\isasymtheta}\ {\isasymin}\ space\ Sp{\isachardoublequoteclose}\isanewline
\ \ \isakeywordTWO{shows}\ {\isachardoublequoteopen}SGD{\isacharunderscore}{\kern0pt}Poisson\ {\isasymtheta}\ {\isacharparenleft}{\kern0pt}xs{\isacharat}{\kern0pt}ys{\isacharparenright}{\kern0pt}\ {\isacharequal}{\kern0pt}\ \ do{\isacharbraceleft}{\kern0pt}{\isasymtheta}{\isacharprime}{\kern0pt}\ {\isasymleftarrow}\ SGD{\isacharunderscore}{\kern0pt}Poisson\ {\isasymtheta}\ xs{\isacharsemicolon}{\kern0pt}\ SGD{\isacharunderscore}{\kern0pt}Poisson\ {\isasymtheta}{\isacharprime}{\kern0pt}\ ys{\isacharbraceright}{\kern0pt}{\isachardoublequoteclose}
\end{isabelle}

We now consider the adjacency based on Hamming distance.
Then adjacent datasets $D_1$ and $D_2$ are decomposed by $D_1 = xs @ [x_1] @ ys$ and $D_2 = xs @ [x_2] @ ys$ where $x_1 \neq x_2$.

By combining the formalization of RDP of a step of loop, the above append law, and 
Lemmas \ref{lem:convexRDP2} we obtain.
\begin{isabelle}
\isakeywordONE{lemma}\isamarkupfalse%
\ Renyi{\isacharunderscore}{\kern0pt}div{\isacharunderscore}{\kern0pt}SGD{\isacharunderscore}{\kern0pt}Poisson{\isacharunderscore}{\kern0pt}conv{\isacharcolon}{\kern0pt}\isanewline
\ \ \isakeywordTWO{assumes}\ adjacent{\isacharcolon}{\kern0pt}\ {\isachardoublequoteopen}xs\ {\isasymin}\ HLists{\isachardoublequoteclose}\ {\isachardoublequoteopen}ys\ {\isasymin}\ HLists{\isachardoublequoteclose}\ {\isachardoublequoteopen}Hdist\ xs\ ys\ {\isacharequal}{\kern0pt}\ {\isadigit{1}}{\isachardoublequoteclose}\ \isanewline
\ \ \ \ \isakeywordTWO{and}\ {\isachardoublequoteopen}{\isasymtheta}\ {\isasymin}\ space\ Sp{\isachardoublequoteclose}\ \isakeywordTWO{and}\ a{\isacharcolon}{\kern0pt}\ {\isachardoublequoteopen}{\isadigit{1}}\ {\isacharless}{\kern0pt}\ {\isasymalpha}{\isachardoublequoteclose}\isanewline
\ \ \isakeywordTWO{shows}\ {\isachardoublequoteopen}Renyi{\isacharunderscore}{\kern0pt}div\ {\isasymalpha}\ {\isacharparenleft}{\kern0pt}SGD{\isacharunderscore}{\kern0pt}Poisson\ {\isasymtheta}\ xs{\isacharparenright}{\kern0pt}\ {\isacharparenleft}{\kern0pt}SGD{\isacharunderscore}{\kern0pt}Poisson\ {\isasymtheta}\ ys{\isacharparenright}{\kern0pt}\ {\isasymle}\ Renyi{\isacharunderscore}{\kern0pt}log{\isacharunderscore}{\kern0pt}part\ {\isasymalpha}\ {\isacharparenleft}{\kern0pt}ennreal\ b\ {\isacharasterisk}{\kern0pt}\ inv{\isacharunderscore}{\kern0pt}Renyi{\isacharunderscore}{\kern0pt}log{\isacharunderscore}{\kern0pt}part\ {\isasymalpha}\ {\isacharparenleft}{\kern0pt}ereal\ {\isacharparenleft}{\kern0pt}{\isadigit{2}}\ {\isacharasterisk}{\kern0pt}\ {\isacharparenleft}{\kern0pt}{\isasymalpha}\ {\isacharasterisk}{\kern0pt}\ C\isactrlsup {\isadigit{2}}{\isacharparenright}{\kern0pt}\ {\isacharslash}{\kern0pt}\ {\isasymsigma}\isactrlsup {\isadigit{2}}{\isacharparenright}{\kern0pt}{\isacharparenright}{\kern0pt}\ {\isacharplus}{\kern0pt}\ ennreal\ {\isacharparenleft}{\kern0pt}{\isadigit{1}}\ {\isacharminus}{\kern0pt}\ b{\isacharparenright}{\kern0pt}{\isacharparenright}{\kern0pt}{\isachardoublequoteclose}
\end{isabelle}
Finally by applying the composition theorem of RDP $T$ times, 
we have RDP of the entire Algorithm \ref{alg:SGD}
(\isa{RDP{\isacharunderscore}{\kern0pt}SGD{\isacharunderscore}{\kern0pt}Poisson{\isacharunderscore}{\kern0pt}iter}).

The formalization of Poisson subsampling costs about 400 lines, and the one of the Algorithm \ref{alg:SGD} costs about 700 lines.

\section{Related Work}
Dong et al. ~ propose the concept of $f$-differential privacy ($f$-DP), which are modeled with trade-off curves between Type I and Type II errors of hypothesis testing for two adjacent datasets~\cite{10.1111/rssb.12454}.
Recently, Fernandes et al. reformulated it as a Galois connection between trade-off curves and two-row information channels, and gave a simpler proof of the composition theorem of $f$-DP~\cite{10.1007/978-3-032-22730-0_22}.
%
There are several related studies for formal verification of probabilistic programs in proof assistants.
In Isabelle/HOL, Eberl et al.~constructed an executable first-order functional probabilistic programming language~\cite{10.1007/978-3-662-46669-8_4}.
Lochbihler formalized protocols with probabilistic oracles~\cite{10.1007/978-3-662-49498-1_20}.
Basin et al.~implemented a framework CryptHOL in for rigorous game-based proofs in cryptography~\cite{crypthol}.
Hirata et al.~developed an extensive Isabelle/HOL library of quasi-Borel spaces~\cite{hirata22,hirata_et_al:LIPIcs.ITP.2023.18,Quasi_Borel_Spaces-AFP}, which is a model for denotational semantics of higher-order probabilistic programming languages with continuous distributions~\cite{HeunenKammarStatonYang2017LICS,Scibior:2017:DVH:3177123.3158148}).
In Rocq, Bagnall and Stewart embedded {\sc MLCert} for formal verification of machine learning algorithms~\cite{Bagnall_Stewart_2019}.
Affeldt et al.~formalized probabilistic programs with continuous random samplings and conditional distributions~\cite{10.1145/3573105.3575691} by formalizing the semantic model based on s-finite kernels~\cite{10.1007/978-3-662-54434-1_32}.

In recent years, Rocq has made significant progress on measure theory:
Lebesgue integrations, Fubini-Tonelli theorem,
Radon-Nikod\'{y}m theorem, and the fundamental theorem of calculus for Lebesgue integration
~\cite{DBLP:journals/jar/BoldoCFMM22,DBLP:conf/fm/BoldoCMMM23,ishiguro2023ppl,affeldt2024jfla}.
In Lean, Tristan et al.~developed an automated measurability prover for probabilistic programs in the continuous setting via reparametrization of the uniform distribution~\cite{tristan2020verification}.
Also, the Lean Mathlib library is actively updated, and it contains the formalization of basic measure theory (see \cite{LeanDoc}).
Recently Luccioli formalized $f$-divergence in Lean 4~\cite{Luccioli_f_divergence}.
The Lean formalization is more elegant than ours, and it covers general generator functions for $\Delta^{\smash{f}}$ and the postprocessing inequality of R\'{e}nyi divergence.
However, the composability of R\'{e}nyi divergence and the approximability of $f$-divergences are not included in the Lean formalization.
Thus, we argue the novelty of our formalization for these results.

We expect that this work will also be useful for users of these proof assistants who want to formalize DP and RDP in the continuous setting.
For example, formalizations of the composability of RDP and conversion laws are useful.
Also, the users might be able to learn from our formalization what fundamental libraries will be required.

\section{Conclusion and Future Work}
In this paper, we give an Isabelle/HOL library for R\'{e}nyi differential privacy in the continuous setting, and give a formalization of the stochastic gradient descent (SGD) mechanism with Gaussian noise and Poisson subsampling.
We have formalized conversion laws from RDP to DP, including Asoodeh et al.'s complicated one.

For a technical basis, we formalize $f$-divergences, which are a class of statistical divergences generated by nonnegative convex functions, and their basic properties.
Our formalization is limited to nonnegative generator functions, but it is enough to formalize R\'{e}nyi divergences.

We formalize the RDP and RDP of Gaussian SGD Poisson subsampling, though the result is limited.
During the trial of formalizations, we found a technical gap in a well-known central lemma for the Gaussian mechanism with subsampling.
We thus took another approach and give a novel (not good but simple) evaluation of subsampling amplification of R\'{e}nyi differential privacy.

\paragraph{Overview}

We show overview of our formalization in Table \ref{table:LOC}.
We use 18,043 lines in total, including comments.
\begin{table}[h]
\centering
\vspace{-0.5em}
\caption{Overview of formalization (Lines of Code).}
\label{table:LOC}
\vspace{-1em}
\begin{tabular}{l|r}
Category & Lines \\\hline\hline
\quad Modification of \isa{Differential\_Privacy}& about 400\\
\quad Miscellaneous lemmas&  862\\
\quad Convex functions& 2,361\\
\quad $f$-divergences for nonnegative generators&  8,178\\
\quad R\'{e}nyi divergence and its composability& 3,203\\ 
\quad Gaussian Mechanism, Gaussian SGD&  2,739\\ \hline\hline
Total &  18,043
\end{tabular}
\vspace{-1em}
\end{table}

\paragraph{Future work}
A topic of future work is to finish the formalization of $f$-divergences for general generator functions.
The formalization of $f$-divergence is not novel because of the existing Lean formalization~\cite{Luccioli_f_divergence}, but it might be meaningful to enrich Isabelle/HOL's library.

For another topic, we plan to improve the formalization of the privacy of Gaussian SGD.
Our current formalization gives a very rough evaluation and is too simple.
We recognize that there is a large room for improvement.
For example, Asoodeh et al. also give a refinement of the law for moment accountants in DP \cite{AbadiCGMMT016}, which is based on Lemma \ref{convRDPtoDP:Asoodeh}~\cite{DBLP:journals/corr/abs-2008-06529}.
For amplification of RDP by subsampling, Lemma \ref{lem:convexRDP2} might give a better evaluation than \cite[Thm. 9]{pmlr-v89-wang19b} for a large subsampling rate $q$ and a small RDP bound of original mechanism, although such a case is not meaningful.
Formalization of \cite[Thm. 9]{pmlr-v89-wang19b} and comparison of them is also possible future work.

We also aim to formalize a variant of Gaussian SGD by Lee and Kifer~\cite{10.1145/3219819.3220076}, it changes the amount of noise according to gradients and privacy budgets already paid.
It might be a good example of the high expressiveness of Isabelle/HOL.
\begin{acks}
Tetsuya Sato was supported by JSPS KAKENHI Grant Number JP24K14893.
\end{acks}
\bibliographystyle{ACM-Reference-Format} 
\bibliography{reference}
\end{document}